\documentclass[twocolumn]{aastex631}

\newcommand{\msh}{MSH 15$-$5{\em 2}}
\newcommand{\kms}{km~s$^{-1}$}
\newcommand{\msol}{$M_{\sun}$}

\newcommand{\ergs}{erg~s$^{-1}$~cm$^{-2}$}
\newcommand{\ergsarcs}{erg~s$^{-1}$~cm$^{-2}$~arcsec$^{-2}$}
\newcommand{\neii}{[Ne\,{\footnotesize II}]}
\newcommand{\neiii}{[Ne\,{\footnotesize III}]}
\newcommand{\ariii}{[Ar\,{\footnotesize III}]}
\newcommand{\siii}{[S\,{\footnotesize III}]}
\newcommand{\siv}{[S\,{\footnotesize IV}]}
\newcommand{\oiv}{[O\,{\footnotesize IV}]}
\newcommand{\feii}{[Fe\,{\footnotesize II}]}
\newcommand{\feiii}{[Fe\,{\footnotesize III}]}
\newcommand{\cloudy}{\textsc{Cloudy}}
\shorttitle{Crystalline Silicate in the SNR \msh}
\shortauthors{Kim, Koo, and Onaka}

\begin{document}

\title{Supernova Ejecta with Crystalline Silicate Dust in the Supernova Remnant \msh}

\correspondingauthor{Hyun-Jeong Kim}
\email{kim.hj@kongju.ac.kr}

\author[0000-0001-9263-3275]{Hyun-Jeong Kim}
\affiliation{Department of Earth Science Education, Kongju National University \\
56 Gongjudaehak-ro, Gongju-si, Chungcheongnam-do 32588, Republic of Korea}
\affiliation{Korea Astronomy and Space Science Institute \\
776 Daedeok-daero, Yuseong-gu, Daejeon 34055, Republic of Korea}

\author[0000-0002-2755-1879]{Bon-Chul Koo}
\affiliation{Department of Physics and Astronomy, Seoul National University \\
1 Gwanak-ro, Gwanak-gu, Seoul 08826, Republic of Korea}

\author[0000-0002-8234-6747]{Takashi Onaka} 
\affiliation{Department of Astronomy, Graduate School of Science, The University of Tokyo \\ 
7-3-1 Hongo, Bunkyo-ku, Tokyo 113-0033, Japan}

\begin{abstract}
IRAS 15099-5856 in the young supernova remnant (SNR) \msh\ is the first and only SNR-associated 
object with crystalline silicate dust detected so far, although its nature and the origin of 
the crystalline silicate are still unclear.
In this paper, we present high-resolution mid-infrared (MIR) imaging observations of the bright central 
compact source IRS1 of IRAS 15099-5856 to study the spatial distributions of gas and dust and 
the analysis of its Spitzer MIR spectrum to explore the origin of IRS1. The MIR images obtained with the
T-ReCS attached on the Gemini South telescope show a complicated, 
inhomogeneous morphology of IRS1 with bright clumps and diffuse emission in \neii\ 12.81~\micron\ and 
Qa 18.30~\micron, which confirms that IRS1 is an extended source externally heated by 
the nearby O star Muzzio 10, a candidate for the binary companion of the progenitor star.
The Spitzer MIR spectrum reveals several ionic emission lines including 
a strong \neii\ 12.81~\micron\ line, but no hydrogen line is detected. 
We model the spectrum using the photoionization code \cloudy\ with varying elemental composition. 
The elemental abundance of IRS1 derived from the model is close to that of SN ejecta with 
depleted hydrogen and enhanced metals, particularly neon, argon, and iron. 
Our results imply that IRS1 originates from the SN ejecta and suggest the possibility of the formation 
of crystalline silicate in newly-formed SN dust. 

\end{abstract}


\section{Introduction}\label{sec:intro}

Silicates are the most common dust species in the interstellar medium (ISM) 
of galaxies. While silicate dust in the ISM of our Galaxy is indicated 
to be mostly amorphous \citep{kemper04, gordon23}, 
a sign of crystalline silicate has also been suggested \citep{do-duy20}.
Crystalline silicates have so far been detected in evolved stars and 
young stars \citep[e.g.,][]{molster99a, malfait99}. Crystallization of amorphous silicate 
grains is suggested to occur in circumstellar disks \citep{molster99b}, 
and several hypotheses of the formation of crystalline silicate have been proposed 
including the radial mixing in the disk and shock waves \citep[e.g.,][references therein]{maaskant15}. 
Crystalline silicate has been also detected in ultraluminous infrared galaxies with 
the crystalline-to-amorphous silicate mass ratios of $\sim$0.1 \citep{spoon06}.
This suggests that supernovae (SNe) may be a source of crystalline silicates, 
leading to a model of evolution of crystalline silicates in galaxies \citep{kemper11}.
\citet{kemper11} propose a model of silicate dust evolution in the galaxy 
that crystalline silicate is formed in SNe and amorphized by cosmic-ray hits, 
suggesting that the crystalline silicate features could be a useful measure of 
the youthfulness of the galaxy.

However, the formation of silicate dust in SNe is uncertain. 
Past observations in near-infrared (NIR) to mid-infrared (MIR) have revealed 
the newly-formed dust in the ejecta of core-collapse SNe although the detected 
dust mass is smaller by more than two orders of magnitude 
than the mass theoretically predicted \citep[e.g,][]{sugerman06, sakon09, szalai11}.
Most of the SN dust have not shown the 10~\micron\ silicate feature, 
suggesting that they are mostly carbonaceous dust, and only 
a few SNe show evidence for the formation of silicate dust \citep{kotak09,shahbandeh23}. 
On the other hand, 
an appreciable amount of dust has been detected in SN 1987A and young supernova remnants (SNRs) 
from far-infrared (FIR) observations \citep[e.g.,][]{matsuura11,chawner20, millard21}, 
but dust composition could not be derived because strong dust features are only present in MIR. 
From MIR spectroscopic observations of SNRs, the presence of carbonaceous dust
is suggested \citep[e.g.,][]{tappe06, andersen11}, and 
some SNRs show an indication of silicate dust including unusual 
non-stoichiometric silicate and metal oxides \citep[e.g.,][]{arendt14,temim17,rho18}. 
But there has thus far not been any observational evidence for the presence of crystalline silicate 
in SNe or SNRs except for one case, \msh\ (G320.4-1.2), for which its association with 
the SN explosion is not yet clearly understood.

\begin{figure}
\epsscale{1.0}
\plotone{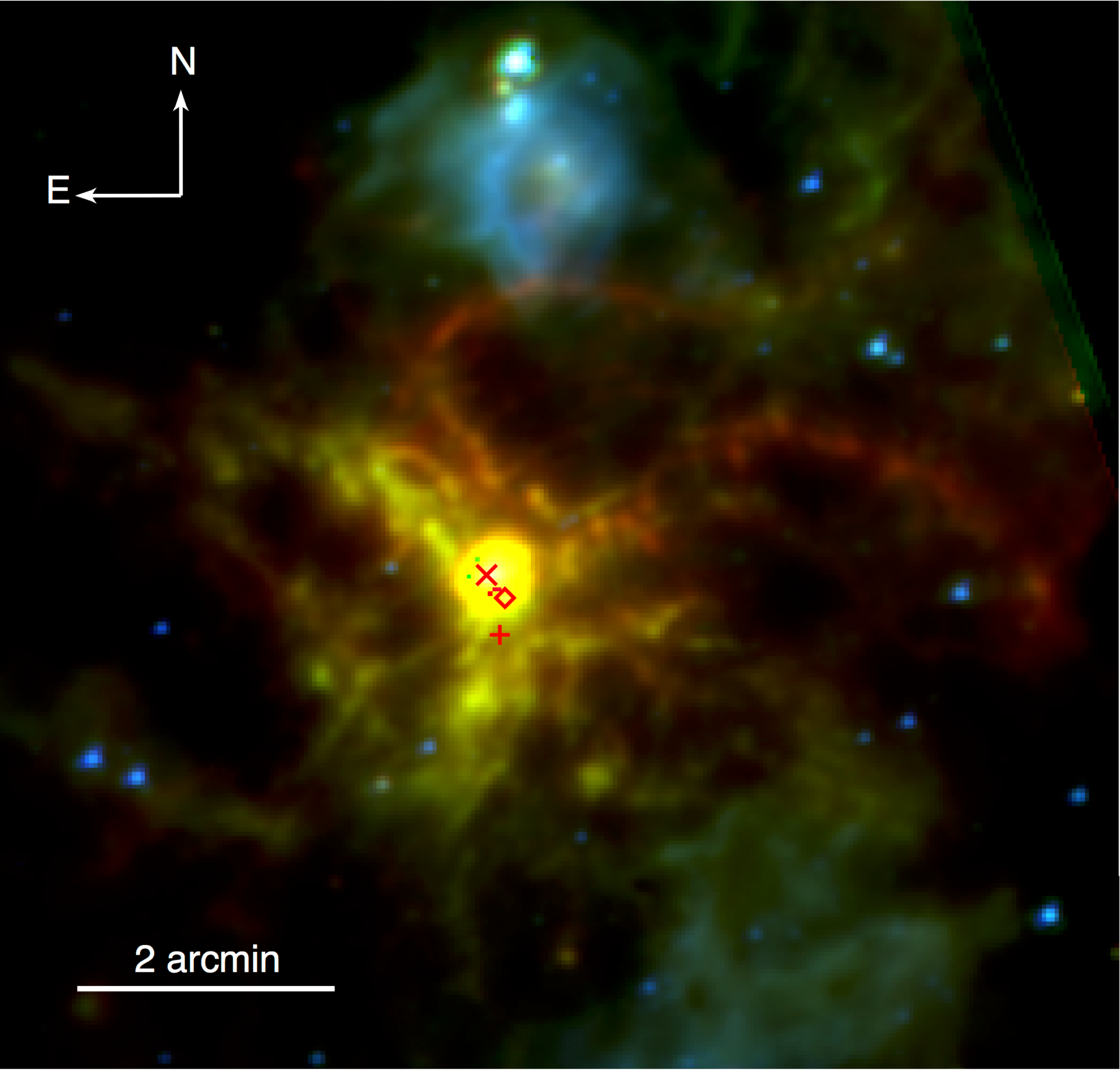}
\caption{Three-color image of IRAS 15099-5856 produced with
AKARI S11 (B, 11~\micron), L15 (G, 15~\micron), 
and L24 (R, 24~\micron) images, which is adopted from 
Figure~1 of \citet{koo11}. The cross ($\times$), diamond, and plus (+) symbols 
present the peak position of IRS1 at 15~\micron, O star Muzzio 10, and 
the pulsar PSR B1509-58, respectively. 
\label{fig:rgb}}
\end{figure}

\msh\ is a young, core-collapse SNR with a complex morphology composed of 
the central pulsar wind nebula and a large radio shell with two distinctive 
components to northwest and southeast separated about 40~pc at 
a distance of 5.2$\pm$1.4~kpc \citep[][but see below]{gaensler99,gaensler02}. 
Although there is no obvious radio counterpart to the pulsar wind nebula, 
the extended X-ray nebula in which the pulsar is embedded appears to coincide 
in position and morphology of the radio components, 
supporting their association \citep{gaensler99}. 
From the large extent of the SNR compared to the young age of 1,700~yrs 
estimated by the central pulsar PSR B1509-58 \citep{seward83}, it was suggested that 
\msh\ is the remnant of Type Ib SN (SN Ib) with a relatively small amount of 
SN ejecta and that the progenitor of the SNR was in a binary system 
with the O star Muzzio 10 (2MASS J15135520-5907516) 
which is $\sim$20$\arcsec$ apart from the pulsar \citep{gaensler99}.
In the SNR, close to the pulsar,
a bright MIR source IRAS 15099-5856 was discovered from
Infrared Astronomical Satellite observations \citep{arendt91}. 
IRAS 15099-5856 is only seen at wavelengths longer than $\sim$10~\micron\ and 
shows a complicated morphology 
with a bright central compact source, a surrounding halo of $\sim$1$\arcmin$ radius 
with knots and spurs, and several extended ($\sim$4$\arcmin$), knotty
arc-like filaments \citep[Figure~\ref{fig:rgb};][]{koo11}. 
In \msh, it is thought that the reverse shock from the surrounding SNR 
has not yet arrived at the pulsar wind nebula \citep{gaensler02}. 
The distinguishable morphology in IR from other wavelengths further suggests 
that IRAS 15099-5856 is not heated by the typical mechanism for the dust 
IR emission in other SNRs, i.e., collisional heating with hot shocked gas, but 
requires an alternate heating source, e.g., the O star Muzzio 10 \citep{arendt91}. 
Hence, IRAS 15099-5856 is likely ``pristine" supernova ejecta material, not 
processed by a reverse shock. Such pristine dusty ejecta has been found 
in Cas A as well, although the heating source in this case is the radiative 
heating from reverse shock \citep{laming20,milisavljevic24}.

\citet{koo11} investigated the central compact source (IRS1) of IRAS 15099-5856 
with the AKARI MIR imaging observations and the Spitzer IRS spectroscopy. 
The absence of emission at short ($\lesssim$10~\micron) wave bands and 
the extended morphology observed in the AKARI images imply that 
IRS1 is an extended source likely heated by a nearby 
O star Muzzio 10 as proposed earlier \citep{arendt91}. 
A unique feature of IRAS 15099-5856 revealed by the Spitzer IRS spectrum is 
the prominent crystalline silicate dust features \citep{koo11}, which has raised an intriguing 
question about the origin of IRAS 15099-5856 because it is the first and only detection of 
crystalline silicate associated with SNRs so far.
\citet{koo11} proposed a scenario that IRS1 is the material from the progenitor of 
the SNR ejected at its final evolutionary stage based on the Spitzer spectrum 
which is well explained by dust models including Mg-rich crystalline silicates and 
the proximity among IRS1, Muzzio 10, and the pulsar PSR B1509-58. 
In this scenario, IRS1 might have survived the SN blast wave as being shielded by Muzzio 10,
the former binary companion star of the progenitor. However, 
the nature of IRAS 15099-5856 and its association with Muzzio 10 and/or 
the central pulsar are still uncertain.

While the proper motion of Muzzio 10 is known as $4.9215 \pm 0.0163$~mas~yr$^{-1}$ 
with position angle (PA) $= 244.2\arcdeg$ (measured from north to east) from 
the Gaia Data Release 3 \citep{gaia,gaiadr3}, 
the proper motion of PSR B1509-58 has only been derived with huge uncertainties. 
\citet{gaensler99} reported one-sigma upper limits on the pulsar's proper motion: 
39~mas~yr$^{-1}$ in right ascension and 52~mas~yr$^{-1}$ in declination.
\citet{leung18} measured the proper motion of $\mu_{\alpha} = 2 \pm 12$~mas~yr$^{-1}$ 
and $\mu_{\delta} = -50 \pm 24$~mas~yr$^{-1}$. 
The proper motion of the pulsar seems to imply that the site of the SN explosion 
about 1,700 yrs ago was $> 1\arcmin$ apart from Muzzio 10, i.e., no association between 
the SN progenitor and Muzzio 10, but more accurate measurements are required to confirm.

The Gaia Data Release 3 \citep{gaia,gaiadr3} also reports the parallax of 0.293$\pm$0.016~mas 
for Muzzio 10. This corresponds to the distance of 3.42$\pm$0.187~kpc and is much closer than 
the known distance to \msh. However, the distance of 5.2$\pm$1.4~kpc to the SNR is the middle 
value of the lower and upper limits from the H\,{\footnotesize I} observation \citep{gaensler99}. 
Since the upper limit of 6.6$\pm$1.4~kpc is just the tangent point at $-$70~\kms, the lower limit 
of 3.8$\pm$0.5~kpc derived from the velocity ($= -55$~\kms) of the H\,{\footnotesize I} absorption 
is only a reliable estimate. This distance is also consistent with the distance 
to the pulsar PSR B1509-58 ($= 4.2\pm0.6$~kpc) derived from the dispersion 
measure \citep[DM $=$ 252.5$\pm$0.3~cm$^{-3}$~pc;][]{cordes02,hobbs04,abdo10}. 
The association of IRAS 15099-5856 and Muzzio 10 seems fairly plausible \citep{koo11}. 
Although there is no direct evidence to suggest the association of these two sources with \msh, 
it seems very unlikely that they are coincidentally located on the same line-of-sight, 
taking account of the possible similar distances. 
In this paper, we assume that the three sources are physically related and assume their 
distance as 3.4~kpc. 

In this paper, we investigate IRS1 of IRAS 15099-5856 as a follow-up of \citet{koo11}.
In Section~\ref{sec:trecsobs}, we present the high-resolution MIR imaging observations of IRS1 
and examine the spatial morphology of IRS1 in detail. In Section~\ref{sec:cloudy}, 
we analyze the Spitzer spectrum of IRS1 using model calculations to derive 
the elemental abundance of gas and investigate dust emission. Particularly, 
we take account of geometry and energy balance to draw a more physically realistic picture of IRS1. 
In Section~\ref{sec:discuss}, we discuss the origin of crystalline silicate in \msh\  
and dust formation in SN ejecta. In Section~\ref{sec:summary}, 
we summarize and conclude our study.

\section{Mid-Infrared Observations of IRAS 15099-5856 IRS1}\label{sec:trecsobs}

\subsection{T-ReCS Observations and Data Reduction}\label{sec:data}

We observed the central compact source IRS1 of IRAS 15099-5856 using 
the Thermal-Region Camera Spectrograph \citep[T-ReCS;][]{telesco98,debuizer05} 
attached on the Gemini South telescope (Program ID: GS-2012A-C-4; PI: Onaka, T.)
on 2012 May 11 UT. 
T-ReCS uses a Raytheon $320 \times 240$ pixel Si:As IBC array, providing a pixel scale 
of $0\farcs089$~pixel$^{-1}$ with a field of view of $28\farcs8 \times 21\farcs6$.
We applied the standard chop-nod technique in order to remove time-variable sky background, 
telescope thermal emission, and the 1/f noise in detector. The chop throw and angle 
were 15$\arcsec$ and 20$\arcdeg$, respectively.
Images were obtained with the Si-6 ($\lambda_0 = 12.33~\micron, \Delta \lambda = 1.18~\micron$),
\neii\ ($\lambda_0 = 12.81~\micron, \Delta \lambda = 0.23~\micron$), 
\neii cont ($\lambda_0 = 13.10~\micron, \Delta \lambda = 0.22~\micron$),
Qa ($\lambda_0 = 18.30~\micron, \Delta \lambda = 1.51~\micron$), and
Qb ($\lambda_0 = 24.56~\micron, \Delta \lambda = 1.92~\micron$) filters, among which 
the Si-6 and \neii cont filters were used to determine the continuum 
baseline of the \neii\ image. For flux calibration, we observed standard stars $\gamma$ Cru and 
$\omega$ Lup (HD 139127) from Cohen standards \citep{cohen99} with the same filters.
The total exposure time of IRS1 was 300 sec for the Qa filter and 900 sec for the others.
The standard stars were observed with the exposure time of 30 sec for all filters.

Data were reduced by using the custom IDL software 
MEFTOOLS version 5.0\footnote{MEFTOOLS was developed and provided by James M. De Buizer via http://www.jim-debuizer.net/research/, but it is no longer available.}.
During the image stacking, bad-frames such as the ones affected by instrumental artifacts 
have been excluded via visual inspection. 
Since IRS1 is an extended source, we need to align the images of different filters. 
In the observed field, the O star Muzzio 10 is the only MIR point source, and it was 
only visible in the Si-6 image. Therefore, we first used Muzzio 10 for the absolute astrometry 
of the Si-6 image. Then, we used standard stars observed in the same 
sequence of filters as IRS1 for correcting the relative astrometry 
among different-filter images with respect to the Si-6 image.
Although this is a rough correction only using one star, 
the peak position of IRS1 in the Qa-band ($\lambda_0 = 18.30~\micron$) is coincident with 
the peak position defined based on the AKARI 15~\micron\ image within $<$0\farcs2.
The seeing estimated from the standard stars is about $0\farcs6$. 
For flux calibration, 
the standard star $\omega$  Lup was used for all the filters except Qb that used 
$\gamma$ Cru. The in-band fluxes of $\omega$  Lup in the Si-6, \neii, \neii cont, and 
Qa filters are 11.381, 10.569, 10.162, and 5.172 Jy, respectively, at airmass of 1, 
similar to the airmass of the IRS1 at the time of observations (1.14--1.4); 
the in-band flux of $\gamma$ Cru in the Qb filter is 157.567 Jy 
at airmass of 1\footnote{https://webarchive.gemini.edu/20210512-sciops-\/-instruments-\/-michelle/find-band-mid-ir-standard-star-fluxes.html}

\subsection{Mid-Infrared Morphology}\label{sec:morphol}

Figure~\ref{fig:trecs} displays the \neii, Qa, \neii cont, Si-6, and Qb images of IRS1 
obtained with T-ReCS and the AKARI S11 ($\lambda_0 = 11~\micron$) image 
with the contours of Qa (green, black) and \neii\ (cyan) overlaid. 
The T-ReCS images were smoothed by a Gaussian kernel with a standard deviation of 1$\arcsec$.
In the AKARI image with low spatial resolution, IRS1 is elliptically extended along east-west direction 
with PA of 110$\arcdeg$. It is also extended in the T-ReCS images 
but shows an irregular morphology with sub structures.
In the Qa image, IRS1 consists of two parts in the east and west. 
The eastern part is extended along northwest-southeast (PA = 144$\arcdeg$) direction
and composed of three bright clumps although the brightest peaks are not well defined.
The western part, on the other hand, is extended along 
northeast-southwest (PA = 50$\arcdeg$) direction and composed of 
a bright compact knot and diffuse emission.
The size of the bright knot obtained by Gaussian fitting is 
about $2\farcs7 \times 1\arcsec$ in FWHM. 
Owing to this bright knot,
the surface brightness of the western part is comparable to the brightness of 
the eastern part (see below) although the western part is about half the size of 
the eastern part.
The \neii\ image is overall similar to the Qa image, but the detailed 
structure is different. The eastern part in \neii\ is extended as large as the eastern part of 
the Qa image, but two bright knots are distinctively shown with 
the size about $1\farcs4 \times 2\farcs7$ and $1\farcs3 \times 1\farcs7$ in FWHM, both of 
which are larger than the seeing ($\sim0\farcs6$) measured from the standard stars. 
The western part in \neii\ is very faint and much smaller than the western part in the Qa image.

\begin{figure*}[!t]
\epsscale{1.0}
\plotone{f2.pdf}
\caption{T-ReCS images of IRS1 compared with 
the AKARI S11 (11~\micron) image. 
The cyan contours on the Qa image are the \neii\ 12.81~\micron\ contours with 
the flux levels of 0.3, 0.55, 0.65, 0.9, 1.2, and 1.5~mJy from the outermost.
The green or black contours on the other images are the Qa 18.30~\micron\ contours
with the flux levels of 0.7, 1.1, 1.5, 2.1, and 2.5~mJy.
On the AKARI S11 image, the cross and diamond symbols present 
the peak position of IRS1 at 15~\micron\ and O star Muzzio 10, respectively.
\label{fig:trecs}}
\end{figure*}

A remarkable feature is that the distributions of Qa and \neii\ emission are not
consistent with each other. Although the whole extent is similar, the bright 
peaks have offsets between the two images as shown by the Qa contours on the \neii\ image
and the \neii\ contours on the Qa image in Figure~\ref{fig:trecs}. 
This discrepancy does not likely come from the inaccurate astrometry.
The relative distances between the peaks are also different in the two images.
The Qa image mostly traces dust distribution since the Qa band covers the wavelengths 
where the amorphous silicate dust continuum and one of the strong features of crystalline 
silicate exists \citep[][see also Section~\ref{sec:spec}]{koo11}. 
Therefore, the Qa and \neii\ images obtained with T-ReCS indicate
not only a complex, inhomogeneous morphology of IRS1 itself but also different
distributions of the gas and dust in IRS1. 

The other T-ReCS images besides Qa and \neii\ do not show any particular emission.
The \neii cont image shows no emission, implying no continuum emission in the \neii\ image.
There is weak continuum around 13~\micron\ 
in the Spitzer IRS spectrum of IRS1 (see Figure~\ref{fig:irsspec}),
but it may be too weak to be detected in the T-ReCS image.
The Si-6 image shows the emission almost identical to \neii. This implies that 
there is no other line except \neii.
While IRS1 is bright at wavelengths longer than 15~\micron\ with 
the spectral energy distribution (SED) peaking at around 30~\micron\ (see Figure~\ref{fig:sed}), 
the Qb image at 24.56~\micron\ in Figure~\ref{fig:trecs} does not detect significant emission 
because of lower sensitivity of the Qb filter.
In the Qb image, some faint emission features are shown inside the Qa contours, but 
they are well below three-sigma (3$\sigma$) 
where $\sigma$ ($\simeq 1.0 \times 10^{-14}$~\ergs\ for Qb) is an rms noise.

Previously, \citet{koo11} suggested that IRS1 is externally heated by Muzzio 10, 
which is 13\farcs7 away from IRS1 to the south, based on 
the non-detection of an embedded point source in optical/NIR imaging and 
the temperature of Muzzio 10 which is appropriate to produce the observed \neii\ line 
luminosity obtained from the Spitzer spectrum (see Section~\ref{sec:cloudy} as well).
The T-ReCS images also do not show any signature of a point source embedded 
in IRS1, confirming the previous prediction. 
A star might be deeply embedded in the bright \neii\ knots, but non-detection of 
any stellar source in \neii cont or Si-6 rules out this possibility.

\subsection{Flux Estimation}\label{sec:flux}

We measured the flux of IRS1 from the Qa and \neii\ images.
Applying aperture photometry, we estimated the flux of the whole source, 
eastern and western parts, and the two 
bright knots in \neii\ as listed in Table~\ref{tbl:flux}. 
The source regions were determined by 
the $3\sigma$ contours,
and the knot regions were determined by the size of the knots. 
The uncertainty in flux measurements is $\lesssim$20\%.
As described earlier, the Qa flux of the western part is smaller than that of 
the eastern part, but the surface brightness of both are comparable.
The \neii\ flux is concentrated in the bright knots.
The bright \neii\ knots have surface brightnesses two times larger than 
that of the eastern region as a whole. 
We also measured the Si-6 flux for the same source region as \neii\ for comparison. 
The Si-6 flux is $1.03 \times 10^{-11}$~\ergs, which is a little larger than the \neii\ flux 
$9.58 \times 10^{-12}$~\ergs\ owing to weak ($\sim 1 \times 10^{-12}$~\ergs) continuum 
at 11.5--13~\micron\ seen in the Spitzer IRS spectrum 
(see Section~\ref{sec:spec} and Figure~\ref{fig:irsspec}).

\begin{deluxetable*}{ccccrc}
\tablecaption{T-ReCS Qa and \neii\ Flux of IRS1\label{tbl:flux}}
\tablewidth{0pt}
\tablehead{
\colhead{Filter} & \colhead{Region} & \colhead{Flux Density} & 
\colhead{Flux} & \colhead{Area} & \colhead{Brightness} \\
  &  & \colhead{(Jy)} & \colhead{($\times 10^{-12}$~\ergs)} &
\colhead{(arcsec$^{2}$)} & \colhead{($\times 10^{-13}$~\ergsarcs)}
}
\startdata
{     }& whole  &   6.85 &   92.6 &  91.17 &   10.2 \\
Qa  & east     &   3.84 &   51.9 &  29.07 &   17.9 \\
{     }& west    &   1.64 &   22.2 &  13.51 &   16.4 \\
\cline{1-6}
{     }& whole   &   2.28 &  9.58 &  71.54 &   1.34 \\
{     }& east      &   1.80 &  7.58 &  32.57 &   2.33 \\
\neii\ & west    &   0.15 &  0.63 &   4.47 &   1.42  \\
{     }& east knot &   0.21 &  0.87 &   1.60 &   5.44  \\
{     }& west knot &   0.34 &  1.45 &   2.62 &   5.53  \\
\enddata
\tablecomments{The uncertainty in flux measurements is $\lesssim$20\%.}
\end{deluxetable*}

We compare the flux measured from the T-ReCS images with the flux estimated 
from the Spitzer IRS spectrum. The Qa flux obtained by using the transmission curve of 
the Qa filter\footnote{https://webarchive.gemini.edu/20210513-sciops-\/-instruments-\/-trecs/filters.html} 
is 10.16~Jy, larger than the Qa flux from the T-ReCS image by a factor of 1.5. 
The \neii\ line flux obtained by Gaussian fitting of the emission line 
(Section~\ref{sec:spec}; Table~\ref{tbl:line}) is $8.21 \times 10^{-12}$~\ergs, 
about 85\% of the \neii\ flux from the T-ReCS image.
The flux differences between the T-ReCS and Spitzer observations can be explained 
by the inhomogeneous morphology of IRS1 and slit-loss correction of 
the Spitzer IRS spectrum as well as in part by sky chopping operation in 
the T-ReCS observations. 
The Spitzer IRS spectrum was obtained with 
two low-resolution modules: the short-low (SL) module covering 5.2--14.5~\micron\ and 
the long-low (LL) module covering 14.0--38.0~\micron.
The two slits perpendicularly placed on IRS1 did not cover the same area 
because of different slit widths \citep[Figure~1 of][]{koo11}, and the SL slit along 
north-south direction with a slit width of $3\farcs7$ only partially covered IRS1, requiring slit-loss correction. 
The slit-loss correction factor was determined by the brightness distribution of the source.
IRS1 was assumed as a Gaussian distribution of $12\arcsec \times 5\arcsec$ in size,
which is very different from the morphology observed in the Qa and \neii\ images.
Since the LL slit width ($10\farcs7$) is larger than the size of IRS1 estimated in 
the Qa image, the Spitzer spectrum possibly includes extended, diffuse emission 
as well that is not detected in the T-ReCS observations, leading to a larger Qa flux
from the spectrum.
We note that \citet{koo11} derived the slit-loss correction factor assuming 
the two-dimensional brightness distribution of IRS1 given by the AKARI images
to match the flux between the AKARI images and Spitzer spectrum,
which results in the Qa and \neii\ flux of 14.3~Jy and $1.2 \times 10^{-11}$~\ergs, respectively.

\section{\cloudy\ Modeling of the Spitzer IRS Spectrum of IRS1}\label{sec:cloudy}

\subsection{Spectral Characteristics}\label{sec:spec}

\begin{figure}
\epsscale{1.1}
\plotone{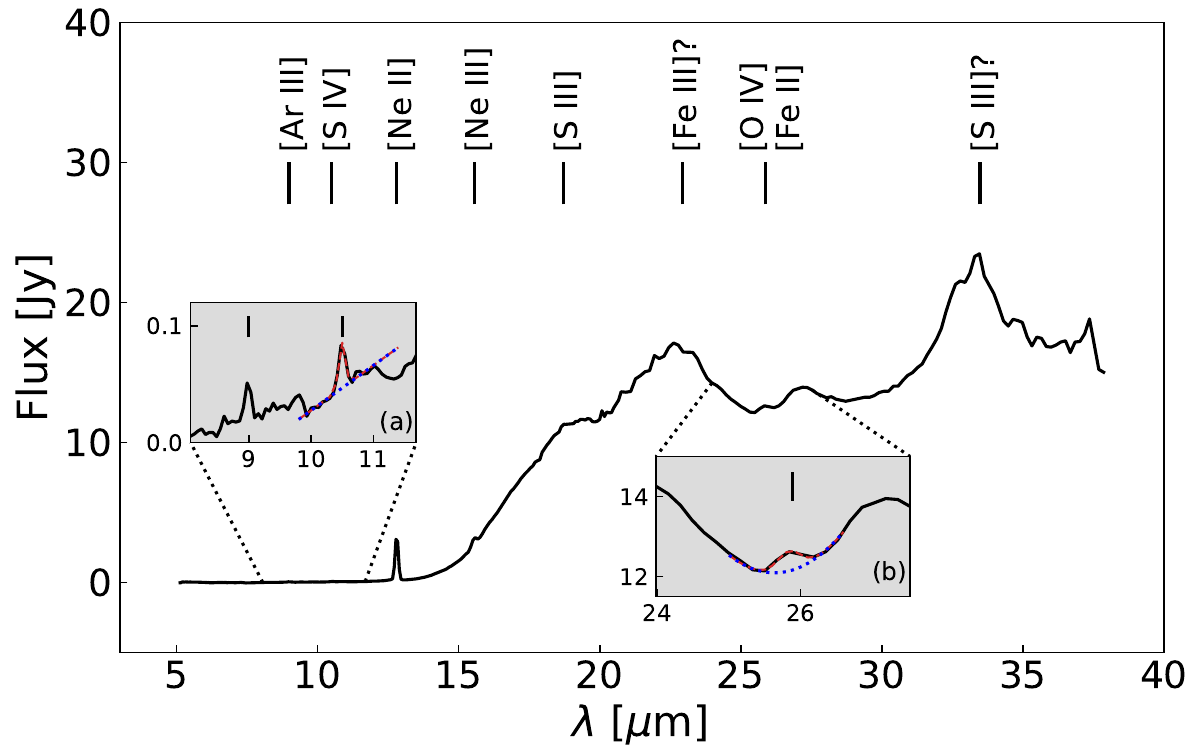}
\caption{Spitzer IRS spectrum of IRS1 \citep{koo11} with the detected lines marked. 
In the insets, the red dashed and blue dotted lines show the fitted Gaussian curve and 
baseline, respectively.
\label{fig:irsspec}}
\end{figure}

The Spitzer IRS spectrum of IRS1 presented in 
Figure~\ref{fig:irsspec} was obtained in 2008 October 3 UT (Program ID: 50495; PI: Koo, B.-C.) 
and examined by \citet{koo11}.   
In the AKARI images from N3 ($\lambda_0 = 3.2~\micron$) to L24 ($\lambda_0 = 24~\micron$), 
IRS1 is only seen at $>$11~\micron. In the Spitzer IRS spectrum likewise, continuum
emission is extremely weak ($<$0.1~Jy) at $\lesssim$13~\micron\ and steeply increases 
to $\sim$20~\micron. The most remarkable features in the spectrum are the strong and relatively 
narrow peaks at 23, 27, and 34~\micron, which are well explained by crystalline silicate dust.
A model of the IRS spectrum produced by modified-blackbody fitting with dust species
including crystalline olivine (Mg$_{1.9}$Fe$_{0.1}$SiO$_{4}$), metal oxides (FeO, MgO), and 
amorphous silicate fairly well reproduces the observed spectrum, providing total dust mass 
of $9 \times 10^{-3}$~\msol\ and dust temperature of $\sim$55--150~K 
at the assumed distance of 4~kpc \citep[see][for details]{koo11}.

The spectrum also shows several ionic emission lines including a strong \neii\ 12.81~\micron\ line 
that was partly discussed in \citet{koo11}. In Figure~\ref{fig:irsspec},
the emission lines are not clearly seen because of the prominent dust features except 
the \neii\ line at 12.81~\micron. We detected \ariii\ 8.99~\micron, \siv\ 10.51~\micron,
\neii\ 12.81~\micron, \neiii\ 15.56~\micron, \siii\ 18.71~\micron, and \oiv\ 25.89/\feii\ 25.99~\micron\ 
lines (Figure~\ref{fig:irsspec}). 
We have not detected [Ar\,{\footnotesize II}] 6.99~\micron\ line, which is generally much 
stronger than \ariii\ 8.99~\micron\ line in SNRs, e.g., RCW 103 and Cas A \citep{oliva99,smith09}. 
The non-detection of [Ar\,{\footnotesize II}] 6.99~\micron\ line, however, is consistent 
with our model where the gas is not shock-ionized but photoionized 
by an O star (see Section~\ref{sec:specm}). 
We also have not detected any hydrogen lines. 
For the detected lines, we measured the line fluxes by the Gaussian fitting using 
the IDL MPFIT package \citep{markwardt09}\footnote{\href{https://pages.physics.wisc.edu/~craigm/idl/fitting.html}{https://pages.physics.wisc.edu/{\char`\~}craigm/idl/fitting.html}}. 
The derived line fluxes are listed in Table~\ref{tbl:line}. 
The flux errors in the table are from the Gaussian fitting and do not include 
the systematic uncertainty of the observed spectrum.
Since the resolving power of the Spitzer IRS LL module\footnote{Spitzer Space Telescope 
Observer's Manual version 8.0, Chapter 7.1.6, issued by the Spitzer Science Center 
(\href{http://ssc.spitzer.caltech.edu}{http://ssc.spitzer.caltech.edu})} between 14 and 21.3~\micron\ is 
given as $R=2.9524\lambda$, i.e., R $\sim$76 at 25.9~\micron, 
two adjacent lines \oiv\ 25.89~\micron\ and \feii\ 25.99~\micron\ are not 
resolved.  
The \siii\ line at 33.48~\micron\ also seems to present in the spectrum, but the line 
flux was not measured because the line is severely blended with the strong dust 
feature at 34~\micron. The FWHM of the emission lines obtained from the Gaussian fitting is 
from 0.11 to 0.39~\micron\ depending on the wavelength. These line widths are comparable to 
the spectral resolving power, implying that the velocity of the ionic lines is not resolved.

\begin{deluxetable}{lcc}
\setlength{\tabcolsep}{0.15in}
\tablecaption{Detected Emission Lines and their Fluxes 
from Spitzer IRS Spectrum of IRS1\label{tbl:line}}
\tablewidth{0pt}
\tablehead{
\colhead{Line} & \colhead{Wavelength} & \colhead{Flux} \\
\colhead{ } & \colhead{\micron} & \colhead{($\times 10^{-12}$ \ergs)}
}
\startdata
\ariii\  &   8.99 &  0.13($\pm$0.02) \\
\siv\    &  10.51 &  0.14($\pm$0.02) \\
\neii\   &  12.81 &  8.21($\pm$0.02) \\
\neiii\  &  15.56 &  0.97($\pm$0.04) \\
\siii\   &  18.71 &  0.51($\pm$0.07) \\
\oiv/\feii\tablenotemark{a}  &  25.89/25.99 &  0.93($\pm$0.07) \\
\enddata
\tablenotetext{a}{\oiv\ and \feii\ lines are not resolved at the spectral resolution of 
the Spitzer IRS LL module.}
\end{deluxetable}

\subsection{Emission Line Ratios}\label{sec:lratio}

\begin{figure}
\epsscale{1.1}
\plotone{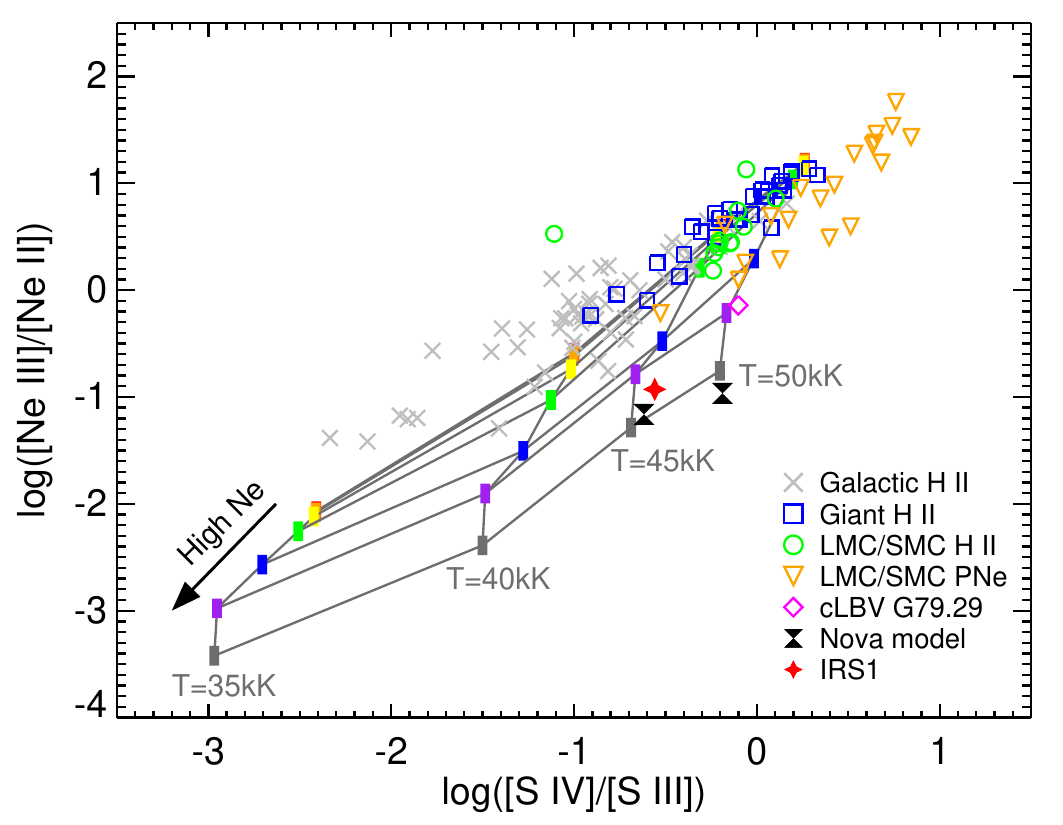} 
\caption{\neiii$_{15.56\micron}$/\neii$_{12.81\micron}$ versus 
\siv$_{10.51\micron}$/\siii$_{18.71\micron}$ line ratio diagram of 
various astronomical objects (gray cross and open symbols), 
\cloudy\ models of novae (black filled hour glass), and IRAS 15099-5856 IRS1 (red star). 
The gray lines are the \cloudy\ model grid produced with hydrogen 
density $n({\rm H})$=100~cm$^{-3}$, heating source temperature 
from 35,000~K to 50,000~K, and ISM abundance with varying neon abundance. 
The small vertical bars in orange, yellow, green, blue, purple, and gray indicate 
neon abundance from $-4.0301$ to $-1.5301$ with an interval of 0.5 in log scale 
relative to hydrogen.
\label{fig:lratio}}
\end{figure}

The IR fine-structure emission lines are frequently used as a diagnostic tool 
in the investigations of gaseous nebulae, ionized regions, or obscured clouds. 
Particularly,
the line ratios of some specific lines have a tight correlation, providing physical conditions 
of the region of interest \citep{dinerstein95,dopita03}. In Figure~\ref{fig:lratio}, we present 
the line ratio diagram \neiii$_{15.56\micron}$/\neii$_{12.81\micron}$ versus 
\siv$_{10.51\micron}$/\siii$_{18.71\micron}$ of various astronomical objects
with the observed line ratios of IRS1.
The line ratios of the objects except novae were obtained from the literature: 
H\,{\footnotesize II} regions in the Galaxy and Large/Small Magellanic Clouds (LMC/SMC) 
from Tables~2 and 5 of \citet{giveon02}; 
giant H\,{\footnotesize II} regions from Table~2 of \citet{lebouteiller08}; 
LMC/SMC planetary nebulae (PNe) from Table~2 of \citet{bernard08}; 
a luminous blue variable candidate (cLBV) G79.29+0.46 from \citet{jimenez10}.
The line ratios of novae were obtained from the model calculations using the one-dimensional 
plasma simulation code \cloudy\footnote{https://www.nublado.org} 
version C13 \citep{ferland13}. \cloudy\ solves the ionization, chemical, and 
thermal state of material exposed to an external radiation field or other heating source, 
and predicts observable quantities such as emission 
and absorption spectra that can be compared with observations.
The nova models were calculated by assuming a blackbody of $T_{\rm eff}=47,000$~K and 
$L = 6.3 \times 10^{36}$~erg~s$^{-1}$ \citep{schwarz07b} and 
by adopting the abundances of the novae V1500 Cygni \citep{ferland78} and 
V838 Her \citep{schwarz07a}, which show enhanced metal abundances (see Table~\ref{tbl:abun}).
We also overlay a \cloudy\ model grid to examine how the line ratios depend on the physical 
parameters. The model grid was produced with hydrogen density $n({\rm H})$=100~cm$^{-3}$, 
heating source temperature from 35,000~K to 50,000~K, and the ISM 
abundance with varying neon abundance from $-4.0301$ (the ISM abundance)
to $-1.5301$ in log scale relative to hydrogen, 
i.e., $n({\rm Ne})$ from $9.33 \times 10^{-3}$~cm$^{-3}$ to 2.95~cm$^{-3}$. 
For the ISM abundance, 
we adopted the protosolar abundance \citep[][see also Table~\ref{tbl:abun}]{asplund09}.
The shapes of the radiation fields (SEDs) of heating sources were adopted from 
the pre-calculated stellar atmospheric models of the Tlusty OB star grids \citep{lanz03,lanz07} 
provided along with the \cloudy\ code from which we selected the main-sequence star models 
at solar metallicity for a given temperature.

It has been known that there is a good correlation 
between \neiii$_{15.56\micron}$/\neii$_{12.81\micron}$ and 
\siv$_{10.51\micron}$/\siii$_{18.71\micron}$ \citep[e.g.,][]{martin02}.
The relation is almost linear, and it suggests that the two line ratios are almost equally 
affected by the hardness of the ionizing radiation. In terms of the stellar $T_{\rm eff}$, 
the observed line ratios of the H\,{\footnotesize II} regions of low to high ionization structures 
can be described with 
$T_{\rm eff}=$ 35,000 K to 50,000 K \citep[Figure~\ref{fig:lratio}; see also Figure 2 of][]{martin02}. 
Figure~\ref{fig:lratio} shows that the relation is in general consistent with the theoretical relation 
expected for H\,{\footnotesize II} regions with the ISM abundance, 
although the majority of the H\,{\footnotesize II} regions with low-ionization structure 
appears to be above the theoretical line. In contrast, the nova models are located well 
below the theoretical relation for the ISM abundance, which is likely due to 
the high abundances of heavy elements. The \cloudy\ model grid, for example, 
demonstrates how the line ratios vary with neon abundance, and it indicates that 
the enhanced neon abundance lowers the \neiii$_{15.56\micron}$/\neii$_{12.81\micron}$ ratio. 
The observed line ratios of IRS1 are similar to those of nova, implying that 
the elemental composition of IRS1 might be similar to that of nova. 
Hence, we adopt the nova abundance as an initial abundance set for our modeling of 
the Spitzer IRS spectrum of IRS1 in Section~\ref{sec:specm}.

We note that there is an issue about the line ratios of IRS1 derived from 
the Spitzer IRS spectrum. 
As described in Section~\ref{sec:flux}, 
IRS1 was observed with two IRS modules with different slit widths 
that covered different parts of IRS1. Since the two lines of each pair in Figure~\ref{fig:lratio} 
(\siv\ and \siii; \neiii\ and \neii) are from different modules, their uncertainty can be large depending 
on the slit-loss correction. 
We compare the line ratios of IRS1 on the empirical relation 
$\rm{ log\,(\neiii /\neii) = 0.81 \times log\,(\siv/\neii)+0.36}$ \citep{groves08} which was derived 
from the archival spectra of a wide range of astrophysical objects from 
nearby H\,{\footnotesize II} regions to ultraluminous infrared galaxies obtained by Spitzer and 
Infrared Space Observatory \citep[ISO;][]{kessler96}. 
With the \siv/\neii\ ratio derived from the same SL module, the \neiii/\neii\ ratio expected
by this relation is 0.09, which is comparable to \neiii/\neii\ $\sim$0.12 derived from
the observed spectrum.
Therefore, we assume that the slit-loss correction is acceptable,
although there is still uncertainty from the brightness distribution between the one
we assumed (i.e., 2D Gaussian distribution) and the real distribution of IRS1.

\subsection{Model Parameters and Assumptions}\label{sec:model}

\begin{figure}
\epsscale{0.7}
\plotone{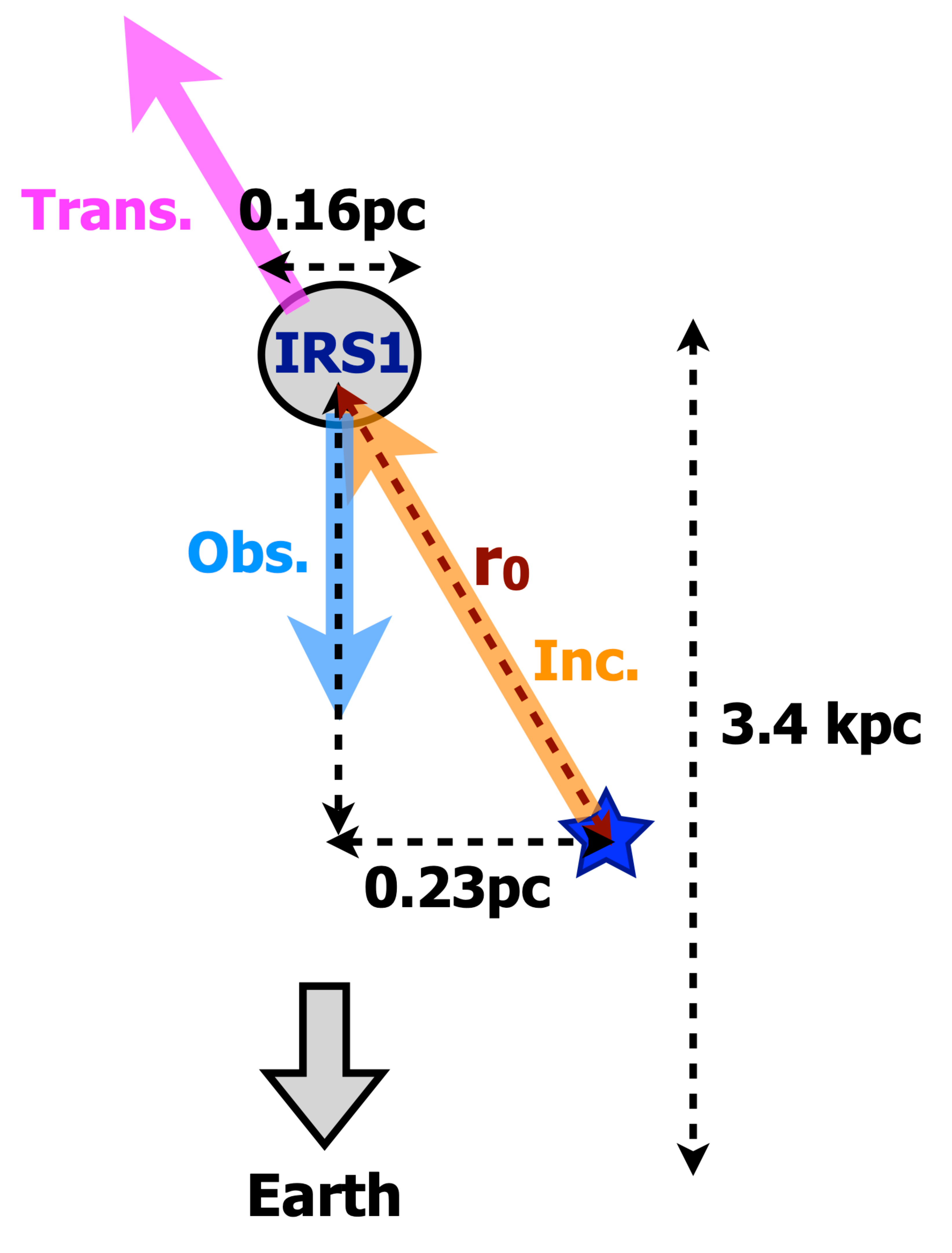}
\caption{Geometry of \cloudy\ model calculations of IRS1. 
The blue star is the heating source Muzzio 10. 
`Inc.' refers to the incident continuum radiation from Muzzio 10.
`Obs.' refers to the emission we observe. This includes 
the backscattered incident radiation of Muzzio 10 and 
the radiation emitted by IRS1 itself.
`Trans.'  refers to the transmitted radiation that includes the attenuated 
incident radiation of Muzzio 10 and the radiation by IRS1.
 In the model calculations, the separation 
between IRS1 and Muzzio 10 ($r_{0}$) has been fixed as 0.45~pc 
(See Section~\ref{sec:specm}).
\label{fig:geo}}
\end{figure}

Previously,
\citet{koo11} modeled the Spitzer IRS spectrum as thermal emission 
from several independent dust components using modified blackbodies. 
Their models well reproduce the observed spectrum, but 
the derived dust temperatures show large differences ranging 
from 55~K to 150~K because they treated dust components independently. 
In this study, we model the IRS spectrum using \cloudy\ to include 
physical process and energy balance. 
We first set the geometry of model calculation. 
The observations imply that IRS1 with 
a size $9\farcs6 \times 5\farcs1$ \citep[from the AKARI image;][]{koo11} or 
0.16~pc $\times$ 0.08~pc at the distance of 3.4~kpc, is externally heated by 
Muzzio 10 separated by $13\farcs7$ or 0.23~pc, in projected distance. 
The separation between IRS1 and Muzzio 10 ($r_{0}$) in principle 
should be treated as a free parameter because it significantly affects 
the radiation absorbed by IRS1 and total dust mass, 
but we fixed it to reduce the number of free parameters 
based on the initial models (See Section~3.4).
We also fixed the thickness of IRS1 as the same as the major axis (0.16~pc) 
of IRS1 on the projected sky.
Figure~\ref{fig:geo} is a schematic figure of the geometry. 
Since the heating source (= Muzzio 10) is outside the cloud (= IRS1),
we adopted a covering factor ($=\Omega / 4\pi$, where $\Omega$ is 
an area of the cloud divided by the distance between the heating source and cloud.) 
to take account into a fraction of the radiation field emitted 
by the heating source that actually strikes the cloud.
In Figure~\ref{fig:geo}, `Obs.' is what we observe, which is the emission 
from the illuminated face of the cloud back into the direction towards 
the heating source. This includes the backscattered incident radiation of Muzzio 10 
and the radiation emitted by IRS1 itself. Meanwhile, the `transmitted (Trans.)' radiation is 
the net emission emerging from the shielded face of the cloud, which includes 
the attenuated incident radiation of Muzzio 10 and the radiation by IRS1. 
If the geometry is reversed, i.e., Muzzio 10 is behind IRS1, 
the transmitted radiation would be what we observe, and it will show the attenuated 
stellar continuum of Muzzio 10 in optical and NIR (see Figure~\ref{fig:sed}).

The heating source in model calculations was fixed as Muzzio 10. 
The spectral type of Muzzio 10 is 
O4.5III(fp) (M. Bessell 2010, private communication) or O5n(f)p \citep{maiz16}. 
Since the luminosity class is uncertain for the latter, we adopted the stellar 
parameters of an O4.5III star \citep{martins05}:
$T_{\rm eff} = 40,500$~K, ${\rm log}~g = 3.71$~cm~s$^{-2}$, and log~$L/L_{\sun}$ =5.76 
or ${\rm log}~Q_{0} =49.52~{\rm s}^{-1}$. For the shape of the radiation field,  
we used the Tlusty O star model at solar metallicity with $T_{\rm eff} = 40,000$~K and 
${\rm log}~g = 3.75$~cm~s$^{-2}$ \citep{lanz03}.  

\begin{figure}
\epsscale{1}
\plotone{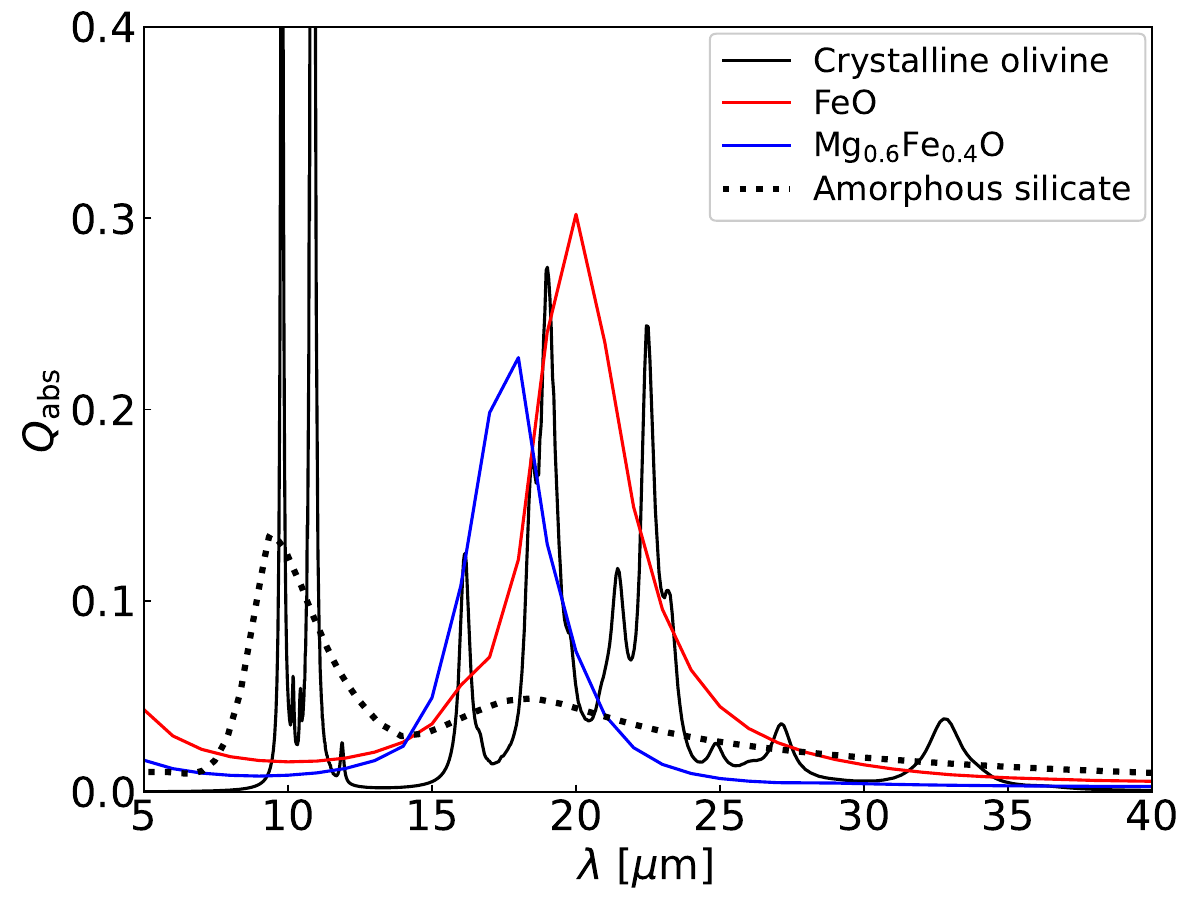}
\caption{Absorption efficiency ($Q_{\rm abs}$) of dust species used in \cloudy\ model 
calculations. The black, red, and blue solid lines represent crystalline 
olivine (Mg$_{1.9}$Fe$_{0.1}$SiO$_{4}$), FeO, and Mg$_{0.6}$Fe$_{0.4}$O, 
respectively. The black dotted line represents amorphous silicate. 
The size of a dust grain is 0.1~\micron\ in radius 
except FeO with a radius of 0.25~\micron.
\label{fig:opac}}
\end{figure}

Dust species were adopted from \citet{koo11}. 
We calculated the absorption/scattering coefficients of each dust from 
their optical constants following the Bohren-Huffman Mie scattering \citep{bohren83} 
and compiled dust opacity files using the grain code in \cloudy. 
The optical constants of dust were adopted from the literature:
crystalline olivine (Mg$_{1.9}$Fe$_{0.1}$SiO$_{4}$) from \citet{fabian01}; 
FeO and Mg$_{0.6}$Fe$_{0.4}$O from \citet{henning95}. 
For amorphous silicate, we used the opacity provided in \cloudy. 
The compiled opacity assumes a spherical dust grain with a size of 
$0.25$~\micron\ for FeO and $0.1$~\micron\ for the others.
Figure~\ref{fig:opac} presents 
the absorption efficiency ($Q_{\rm abs}$) of each dust species.

\subsection{\cloudy\ Models of IRS1}\label{sec:specm}

With the assumed geometry, heating source of an O4.5III star, and 
adopted dust species, we modeled IRS1 using \cloudy\ to derive 
its physical and chemical characteristics. 
Since a large number of free parameters are involved in \cloudy\ calculations, 
we constrained some parameters from the observations. 
There is no hydrogen line emission detected in the Spitzer spectrum or 
optical/NIR images. This requires very low hydrogen density, so we have 
assumed that hydrogen is depleted in IRS1 and 
fixed the hydrogen density as $n({\rm H}) \sim 10^{-5}$~cm$^{-3}$. 
In \cloudy, the initial abundance can be adopted from
the stored abundance sets, e.g., H\,{\footnotesize II} regions, 
general ISM, novae, and PNe. 
Based on the line ratio diagram (Figure~\ref{fig:lratio}, Section~\ref{sec:lratio}), 
we adopted the gas-phase abundance of the nova V1500 Cyg listed in 
the third column of Table~\ref{tbl:abun} but with little hydrogen 
as an initial gas-phase abundance set and adjusted the amounts of the elements. 
The elements not listed in Table~\ref{tbl:abun} were not included in the calculations. 
Grains were included by specifying dust opacity files (Section~\ref{sec:model}).

\begin{deluxetable*}{cccccc}
\setlength{\tabcolsep}{0.1in}
\tablecaption{Abundances used in the \cloudy\ Modeling 
\label{tbl:abun}}
\tablewidth{0pt}
\tablehead{
\colhead{Atom} & \colhead{ISM\tablenotemark{a}} & 
\colhead{Nova (V1500 Cyg)\tablenotemark{b}} & 
\colhead{Nova (V838 Her)\tablenotemark{c}} & \multicolumn{2}{c}{IRS1 Model} \\
\cline{5-6}
 & & & & \colhead{Gas Phase} & \colhead{Grain}
}
\startdata
H   & 1.00E+02 & 3.16E+07 & 3.16E+07 & 1.26E-05 & \nodata \\
He  & 9.55E+00 & 3.09E+06 & 4.47E+06 & 1.23E-06 & \nodata \\
C   & 2.95E-02 & 2.95E+04 & 6.03E+04 & 5.29E-01 & \nodata \\
N   & 7.41E-03 ($-$0.60) & 3.09E+05 (1.02) & 7.41E+04 (0.09) & 5.51E-04 ($-$2.98) & \nodata \\
O   & 5.37E-02 (0.26) & 5.37E+05 (1.26) & 2.82E+04 ($-$0.33) & 1.91E+00 (0.56) & 6.30E-01 \\
Ne  & 9.33E-03 ($-$0.50) & 6.46E+04 (0.34) & 1.91E+05 (0.50) & 3.25E+01 (1.79) & \nodata \\
Mg  & 4.37E-03 ($-$0.83) & 1.20E+03 ($-$1.39) & 1.58E+03 ($-$1.58) & 2.14E-02 ($-$1.39) & 1.80E-01 \\
Si  & 3.55E-03 ($-$0.92) & 1.12E+03 ($-$1.42) & 2.04E+03 ($-$1.47) & 2.00E-02 ($-$1.42) & 1.43E-01 \\
S   & 1.45E-03 ($-$1.31) & 5.13E+02 ($-$1.76) & 6.92E+03 ($-$0.94) & 9.76E-02 ($-$0.73) & \nodata \\
Cl  & 1.86E-05 ($-$3.20) & 5.89E+00 ($-$3.70) & 5.89E+00 ($-$4.01) & 1.06E-04 ($-$3.70) & \nodata \\
Ar  & 2.75E-04 ($-$2.03) & 1.15E+02 ($-$2.41) & 1.15E+02 ($-$2.72) & 1.86E-01 ($-$0.45) & \nodata \\
Fe  & 3.47E-03 ($-$0.93) & 1.48E+03 ($-$1.30) & 8.51E+03 ($-$0.85) & 1.19E+01 (1.35) & 1.64E-01 \\
\enddata
\tablecomments{
The abundance is the absolute number density (cm$^{-3}$) applied to the \cloudy\ models. 
All the abundances except IRS1 Model are the gas-phase abundance. 
The numbers in parentheses are the number density of the element 
relative to carbon in log scale, which shows the relative abundances among metals.}
\tablenotetext{a}{For ISM, the protosolar abundance \citep{asplund09} is adopted.}
\tablenotetext{b}{The nova abundance stored in \cloudy\ derived by \citet{ferland78} 
for V1500 Cygni.}
\tablenotetext{c}{The nova abundance derived for V838 Her \citep{schwarz07a}.}
\end{deluxetable*}

\begin{figure}[b!]
\epsscale{1.05}
\plotone{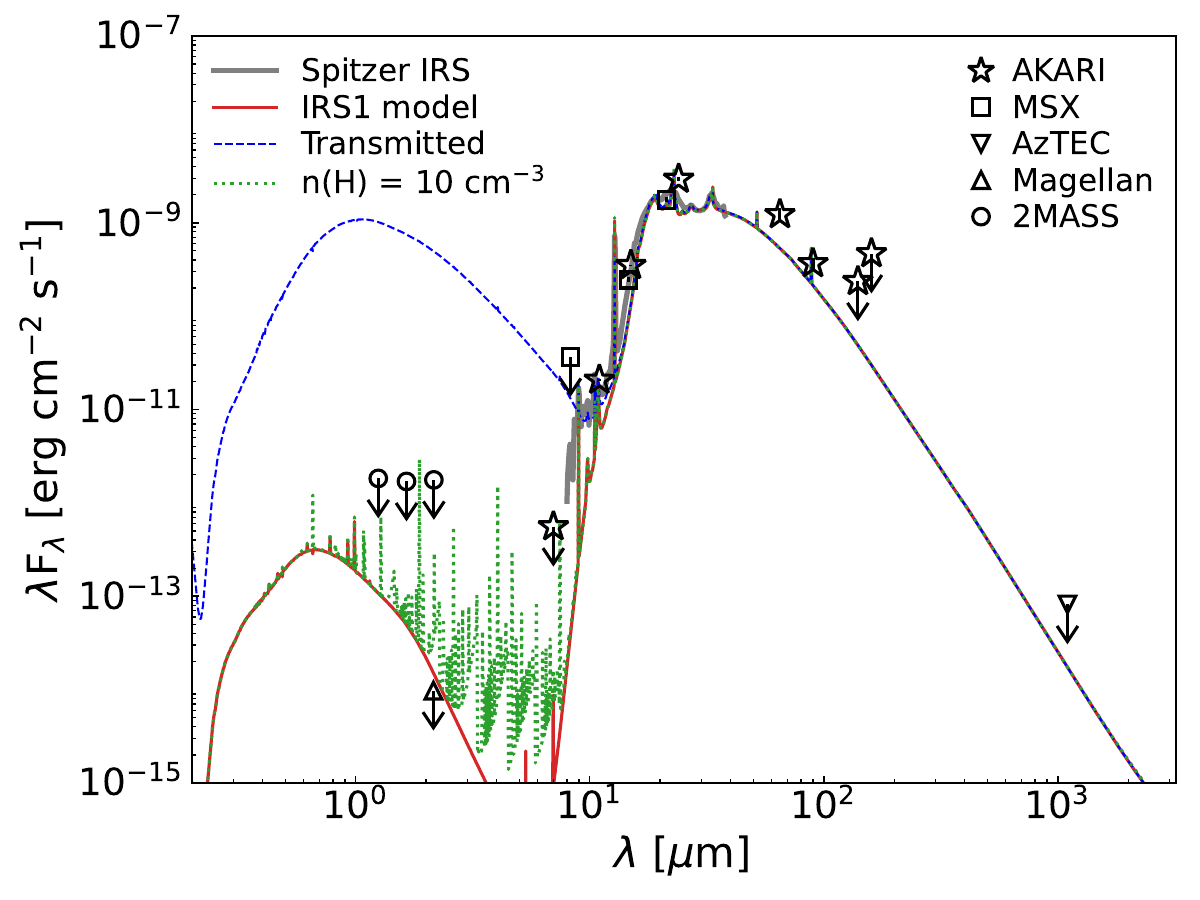}
\caption{SEDs of IRS1 from the observations and \cloudy\ models.
Open symbols are the broadband fluxes from various observations as labeled 
in the legend \citep{koo11}. The symbols with a down arrow represent upper limits. 
The red line is the SED of the reference model of IRS1 that is expected 
to be observed. The blue-dashed line is also the reference model but 
presents the transmitted emission (Section~\ref{sec:model}). 
The green-dotted line shows the model with some amount of 
hydrogen ($n({\rm H})=10$~cm$^{-3}$) for the comparison with 
the reference model with depleted hydrogen ($n({\rm H}) \sim 10^{-5}$~cm$^{-3}$). 
The Spitzer IRS spectrum of IRS1 is also displayed by a thick gray line.
\label{fig:sed}}
\end{figure}

We first determine the scale factor that is applied to all the metals (the elements heavier 
than helium) and grains to fit the observed MIR flux level. 
With the hydrogen density ${\rm log}\,n({\rm H}) = -4.9$, we obtained a scale factor 
of $10^{7.65}$ relative to hydrogen. We also found that $r_{0}=0.45$~pc reasonably well fits 
the observed flux, so we fixed $r_{0}$ as 0.45~pc. 
Then, the abundance of each element is further adjusted to match the observed flux. 
Figure~\ref{fig:sed} shows the Spitzer IRS spectrum of IRS1 with broadband observations. 
In the figure, the flux of the Spitzer spectrum is about 30\% smaller than 
the AKARI 11, 15, and 24~\micron\ broadband fluxes because of the assumed profile 
in the slit-loss correction (Section~\ref{sec:flux}). The AKARI fluxes were derived by 
the aperture photometry with a circle of 25$\arcsec$ radius \citep{koo11}, 
so the aperture contains extended diffuse emission, leading 
to larger fluxes. We fit the \cloudy\ model to the flux of the Spitzer spectrum. 
After roughly fitting the MIR flux level, 
we searched for a model that reproduces the observed dust features by changing 
the amount of four dust species: crystalline olivine, 
FeO, Mg$_{0.6}$Fe$_{0.4}$O, and silicate. With the fixed dust abundance, 
we then determined the gas abundance that explains the observed line flux derived 
in Section~\ref{sec:spec} by changing the densities of six elements 
involved in the formation of the observed emission lines: nitrogen, oxygen, neon, 
sulphur, argon, and iron. These six ions do not independently act but are tightly correlated 
to each other. For example, 
the increased neon abundance does not always lead to stronger neon lines, or
the increased iron abundance strengthens the \neii\ line as well as iron lines 
but not the \neiii\ line.
Since hydrogen is depleted in IRS1, the heating process mostly depends on 
photoelectric heating by dust and heavy elements rather than photoionization by hydrogen; 
thus, changes of the metal and dust abundances affect the heating and cooling processes, 
complicating the model calculations.
With a large number of free parameters and limited observational data,
it is improbable to find the only model that perfectly fits IRS1.
Instead, we intend to present a reference model that reasonably 
well explains the observations and to discuss some parameters 
that affect the modeling results.
The final abundances of the reference model in gas phase and those depleted 
onto grains are presented in the fifth and sixth columns of Table~\ref{tbl:abun}. 
For the gas-phase, we also present the reference metal abundances to carbon 
rather than hydrogen which is depleted in IRS1 for the comparison 
between IRS1 and others.

\begin{figure}[t!]
\epsscale{1.1}
\plotone{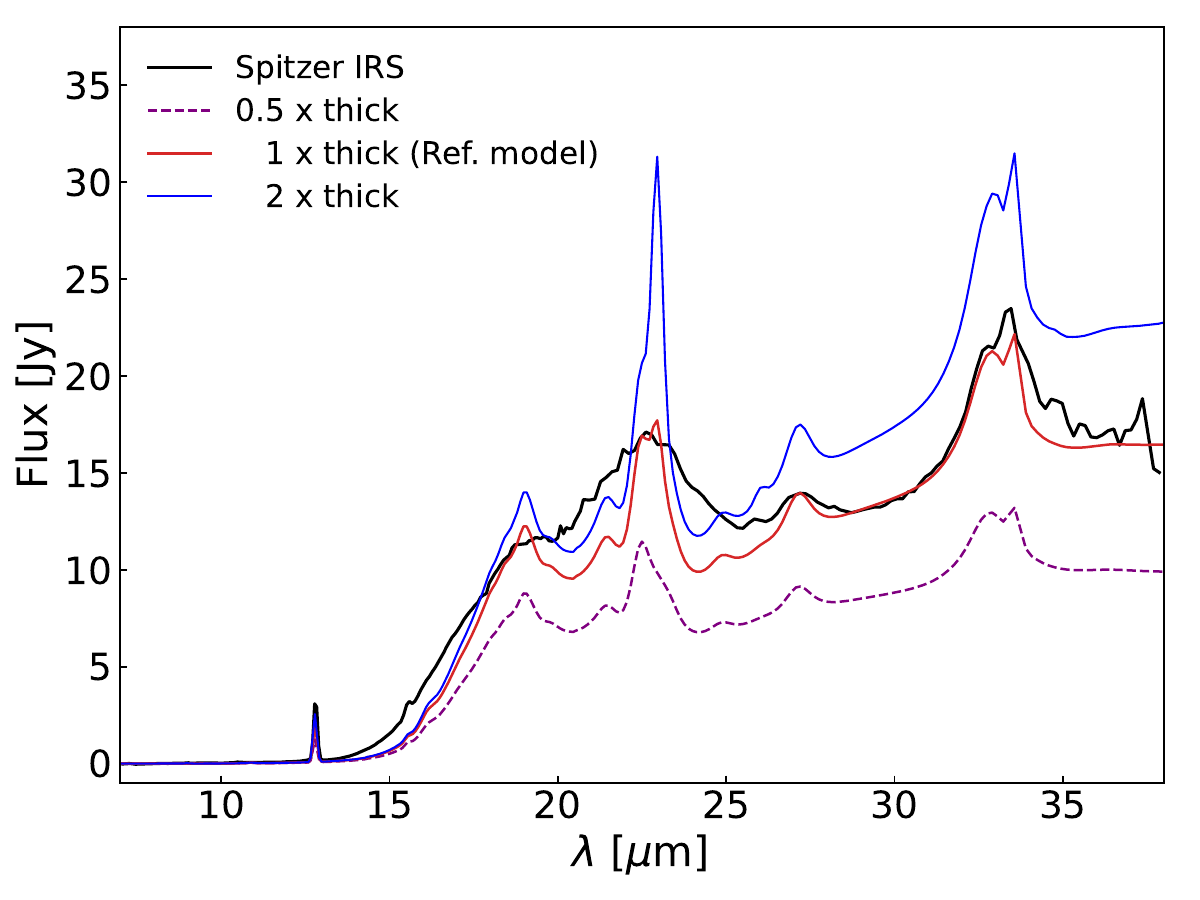}
\caption{Spitzer IRS spectrum (black) of IRS1 and \cloudy\ models with different thickness.
The red solid line is the reference model that fits fairly well the Spitzer IRS spectrum 
in both dust features and line intensities. The purple dashed and blue solid lines are 
the models calculated with the same parameters as the reference model but 
with a half and twice of the thickness, respectively. 
The model spectra were smoothed to the spectral resolution of the Spitzer IRS modules.
\label{fig:spec}}
\end{figure}

In Figures~\ref{fig:sed} and \ref{fig:spec}, we present the reference model of 
IRS1 in red color that fits fairly well the Spitzer IRS spectrum in both dust features 
and line intensities. 
The model was reddened by the column density of $9.5~\times ~10^{21}$~cm$^{-2}$ 
with $R_{\rm V} = 3.1$ \citep{gaensler02,koo11}.
In Figure~\ref{fig:sed}, the red curve presents the SED of the reference model 
expected to be observed for the assumed geometry (Figure~\ref{fig:geo}). 
The model SED again confirms that IRS1 is externally heated. 
If a heating source is embedded in IRS1, we would observe an SED similar to 
the transmitted emission displayed by a blue-dashed curve in Figure~\ref{fig:sed} 
that exhibits strong attenuated incident stellar continuum in optical and NIR. 
An embedded star with low temperature may suppress the optical/NIR flux, but 
such temperature is too low to produce the observed \neii\ line. 

In Figure~\ref{fig:spec}, the model spectrum was smoothed to the spectral resolution of 
the Spitzer IRS modules. The model reproduces the observed dust features 
at 18, 23, 27, and 34~\micron\ but with narrower width for the 23~\micron\ peak. 
The continuum slope from 15 to 20~\micron\ in the model is 
also steeper than observed. These differences can be explained by the sizes and 
shapes of dust grains. We have assumed a spherical dust grain 
of 0.1~\micron\ (or 0.25~\micron\ for FeO) because of the limited availability of 
the optical constants or opacities of dust in MIR, but dust properties in fact 
highly depend on both dust size and shape \citep[e.g.,][]{koike89,koike10,min15}. 
For example, the 10~\micron\ silicate feature becomes broader and is shifted to 
longer wavelength as the grain shape deviates from a perfect sphere. 
The prominent features of forsterite (Mg-rich crystalline silicate) are also significantly 
suppressed for the dust grains with larger size \citep[Figure 5 of][]{min15} 
or elliptical shape \citep[Figure 9 of][]{koike10}. This implies that 
the model of the IRS spectrum can be improved by using elliptical and/or larger 
dust grains.
We note that the dust models in \citet{koo11} 
that used a continuous distribution of ellipsoids (CDE) for FeO 
give a better fit for the steep continuum shape at 15--20~\micron, 
but their models produced by independent dust components
result in the temperature of 90--150~K for FeO and 
Mg$_{0.6}$Fe$_{0.4}$O (or MgO), which is much higher than the temperature of 
the other dust components around 55~K.

\begin{deluxetable}{lcc}
\tablecaption{Dust Parameters of IRS1 from the Reference Model 
\label{tbl:dustm}}
\tablewidth{0pt}
\tablehead{
\colhead{Component} & \colhead{Mass} & \colhead{Temperature}  \\
\colhead{ } & \colhead{($10^{-3}~M_{\sun}$) } &  \colhead{(K)}
}
\startdata
Crystalline olivine (Mg$_{1.9}$Fe$_{0.1}$SiO$_{4}$) & 0.05 & 79 \\
FeO  & 0.01  & 67 \\
Mg$_{0.6}$Fe$_{0.4}$O  & 0.21 & 78 \\
Amorphous silicate & 1.74 & 71 \\
\cline{1-3}
Total & 2.02 & \nodata \\
\enddata
\end{deluxetable}

Table~\ref{tbl:dustm} presents dust mass and temperature obtained from the reference model. 
Total dust mass is $2.02 \times 10^{-3}$~\msol. The derived dust-to-gas ratio is rather 
high with 0.02. Most dust mass is contributed by amorphous silicate. The contribution from 
crystalline olivine is small but indispensable to fit the observed dust features. The total dust mass 
is $\sim$31\% of the dust mass of $9 \times 10^{-3} d_{4}^{2}$~\msol\ 
(i.e., $6.5 \times 10^{-3}$~\msol\ when scaled to $d = 3.4$~kpc) 
derived from the modified-blackbody fit \citep{koo11}, 
likely due to the constrained geometry. 
Total dust mass highly depends on the geometry such as $r_{0}$ 
or the thickness of IRS1 both of which were fixed in our calculations. 
If we increase $r_{0}$, the dust mass becomes larger. For example, 
a model with $r_{0} = 0.52$~pc gives more than twice the dust mass of 
the reference model but with dust temperatures about 12--15~K lower 
than presented in Table~\ref{tbl:dustm}. 
The models with larger $r_{0}$ thus tend to predict larger fluxes at long ($\gtrsim 100~\micron$) 
wavelengths and make the SED fits worse in FIR. Comparing to \citet{koo11}, 
the relative fraction and temperature of the individual dust species are also different.
This may come from the differences in dust opacity.
While the optical constants of each dust were adopted from the same literature, 
the calculations of dust absorption coefficients are different likely due to 
the assumed shapes and sizes of dust grain, resulting in slightly different dust properties.
For example, 
the strength of two peaks at 18 \micron\ and 23~\micron\ of 
the absorption coefficient of crystalline olivine 
are comparable in our calculations (Figure~~\ref{fig:opac}), while 
the 23~\micron\ peak is only 60\% relative to the 18~\micron\ peak in \citet{koo11}, requiring 
larger mass and lower temperature to fit the observed 23~\micron\ feature.
For FeO, we assume a spherical grain of $0.25$~\micron, whereas \citet{koo11} assumed 
a CDE with a size of 0.1~\micron\ which shows 
the weaker, broader, and asymmetric peak of the absorption coefficient.

\begin{deluxetable*}{ccccccccc}
\tabletypesize{\scriptsize}
\tablecaption{Line Fluxes Predicted from the \cloudy\ Models and Relative Intensities to the Observations 
\label{tbl:linem}}
\tablewidth{-2pt}
\tablehead{
\colhead{Model\tablenotemark{a}} & \colhead{\ariii\ } & \colhead{\siv\ } &
\colhead{\neii\ } & \colhead{\neiii\ } &
\colhead{\siii\ } & \colhead{\feiii \tablenotemark{b} } &
\colhead{\feii\ } & \colhead{\siii \tablenotemark{b}} \\
\colhead{($\times$ thick) } & \colhead{ 8.99~\micron\ } & \colhead{10.51~\micron\ } &
\colhead{12.81~\micron\ } & \colhead{15.56~\micron\ } &
\colhead{18.71~\micron\ } & \colhead{22.93~\micron\ } &
\colhead{25.99~\micron\ } & \colhead{33.48~\micron\ }
}
\startdata
0.5 & 1.01E-13 (0.76) & 9.35E-14 (0.66) & 5.61E-12 (0.68) & 7.63E-13 (0.79) & 
3.43E-13 (0.67) & 2.02E-12 & 5.87E-15 (0.01) & 3.73E-12 \\
1.0 & 1.32E-13 (0.99) & 9.83E-14 (0.70) & 7.19E-12 (0.88) & 7.69E-13 (0.79) & 
5.75E-13 (1.13) & 1.69E-11 & 1.40E-13 (0.15) & 6.82E-12 \\
2.0 & 1.53E-13 (1.15) & 1.00E-13 (0.71) & 8.48E-12 (1.03) & 7.72E-13 (0.80) & 
8.93E-13 (1.75) & 6.29E-11 & 1.95E-12 (2.09) & 1.22E-11 \\
\enddata
\tablecomments{Line flux is in \ergs\ and reddened by 
the column density $9.5~\times ~10^{21}$~cm$^{-2}$ with $R_{\rm V} = 3.1$ \citep{koo11}. 
The numbers in parentheses are the relative intensities 
with respect to the observed fluxes.}
\tablenotetext{a}{A scale factor applied to the cloud thickness.}
\tablenotetext{b}{Not seen in the Spitzer IRS spectrum, but likely blended with 
dust features.}
\end{deluxetable*}

The \cloudy\ model also reproduces several ionic lines observed in the IRS spectrum. 
In Table~\ref{tbl:abun}, the reference model indicates that 
neon, argon, and iron are enhanced in IRS1. 
As we pointed out earlier, the reference model is not the only model 
that can explain IRS1, but the overall trend in abundance 
should not be very different
in order to produce the observed lines.
The predicted lines and their fluxes are presented in Table~\ref{tbl:linem} 
with the relative strength to the observed line fluxes. 
The predicted line fluxes mostly agree with the observations within $\lesssim$20\% 
except the \siv$_{10.51\micron}$ and \feii$_{25.99\micron}$ lines. 
For the \siv$_{10.51\micron}$ line, the observed line flux could have been 
overestimated. As seen in the insets of Figure~\ref{fig:irsspec}, 
the continuum at 10--11~\micron\ is noisy and shows a discontinuity 
with a steeper slope compared to the wavelengths $<$10~\micron.  
Considering the predicted line flux of \siii$_{18.71\micron}$ well 
matched to the observation, the overestimation of the line flux 
of \siv$_{10.51\micron}$ by 10--15\% due to a poor baseline is plausible. 
The flux of the \feii$_{25.99\micron}$ line could have also been 
overestimated because the line is located between dust features 
where the baseline is not reliably defined, but the only 15\% of the predicted 
to observed flux ratio requires another reason besides the overestimation of 
the observed flux (see below). The \cloudy\ model predicts 
the \siii\ line at 33.48~\micron\ and \feiii\ line at 22.93~\micron\ as well. 
While these two lines are predicted to be strong, they are blended with 
the dust features of crystalline olivine at 34~\micron\ and 23~\micron, respectively, 
and so are hardly detectable in the IRS spectrum. In the model spectrum, 
after we smoothed it to the Spitzer IRS resolution, the \siii$_{33.48\micron}$ and 
the \feiii$_{22.93\micron}$ lines have become hidden under 
the dust features (Figures~\ref{fig:irsspec} and \ref{fig:spec}).

\begin{figure*}
\epsscale{1.0}
\plotone{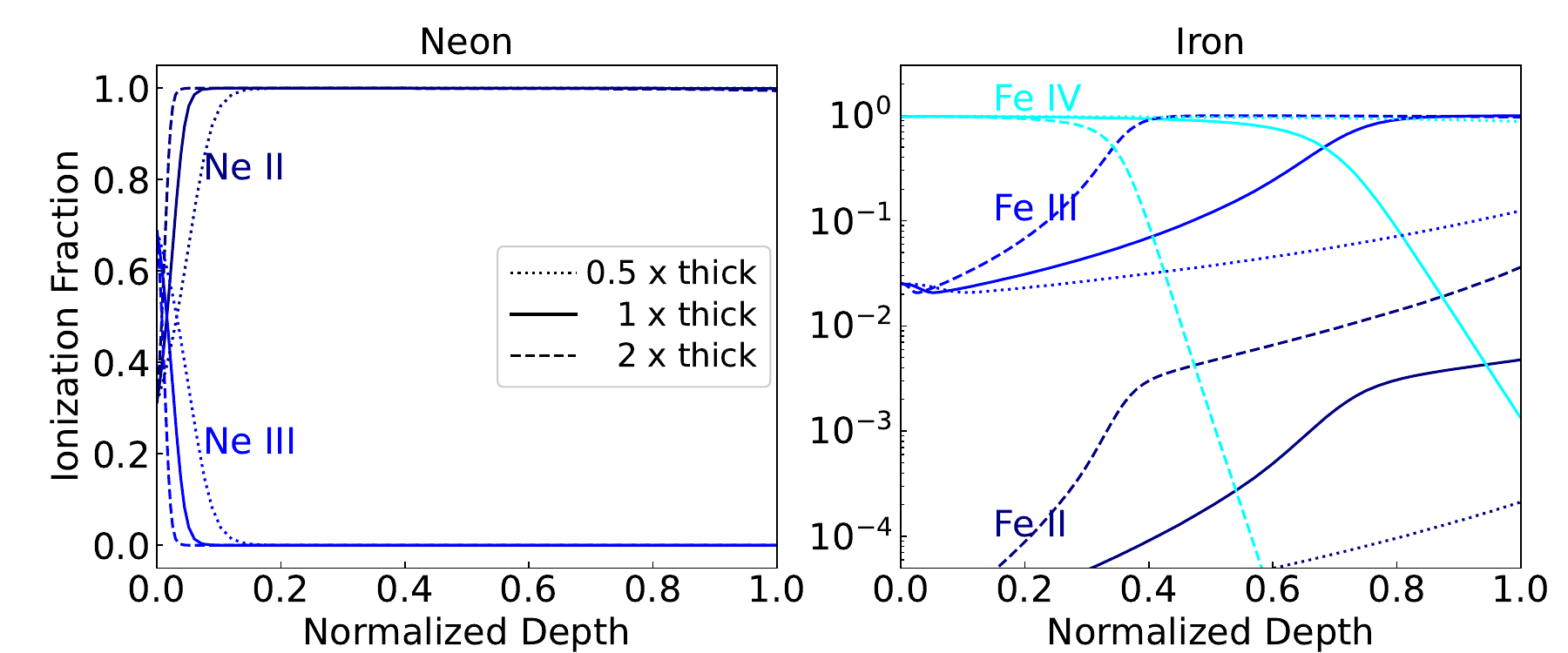}
\caption{
Ionization fraction of neon (left) and iron (right) from the \cloudy\ models of IRS1, 
which is defined as the number density of each ion of an element 
among the total number density of the element.
The solid lines are the reference model; the dashed and dotted lines are the models 
with twice and a half of the thickness, respectively. 
Ions are presented by different colors: navy (Ne\,{\footnotesize II}, Fe\,{\footnotesize II}),
blue (Ne\,{\footnotesize III}, Fe\,{\footnotesize III}), and cyan (Fe\,{\footnotesize IV}).
\label{fig:ion}}
\end{figure*}

As pointed out, the \feii$_{25.99\micron}$ line predicted from 
the model is much weaker than observed. 
While we noted that the line detected at $\sim$25.9~\micron\ could be 
either \oiv$_{25.89\micron}$ or \feii$_{25.99\micron}$, 
the \cloudy\ model only predicts \feii$_{25.99\micron}$.
We also believe that \feii\ is more plausible because 
the UV radiation of an O4.5III star is not hard enough to ionize O\,{\footnotesize III} 
to O\,{\footnotesize IV} for which the ionization potential is 54.9~eV.
The reason for the weaker \feii\ line in the model is because 
most iron is in Fe\,{\footnotesize III} or Fe\,{\footnotesize IV} as shown 
in Figure~\ref{fig:ion} \citep[see also][]{bautista98}. 
The \feii\ line becomes stronger when the thickness of IRS1 increases. 
In Figure~\ref{fig:spec} and Table~\ref{tbl:linem}, 
we present two models produced by the same parameters as the reference model 
except for the cloud thickness: the models with the thickness twice and a half of 
the thickness of the reference model.
In Table~\ref{tbl:linem}, 
the \feii$_{25.99\micron}$ and \feiii$_{22.93\micron}$ lines are the only lines 
that significantly vary by thickness. This behavior is also seen in Figure~\ref{fig:ion} 
which presents the ionization fraction of neon and iron by depth. 
In the figure, depth is normalized and is zero at the closest side to 
the heating source, where the temperature is the highest. 
The ionization fraction is defined as the number density of each ion of an element 
among the total number density of the element.
Figure~\ref{fig:ion} shows that the ionization fraction of iron is more sensitive 
to the thickness compared to neon. For example,
all of the neon is in Ne\,{\footnotesize II} from the very inside ($\lesssim$ 0.15 of the depth) 
regardless of the cloud thickness. In contrast, the fraction of iron ions changes through 
the whole depth depending on the cloud thickness. 
Therefore, the model of IRS1 could be improved by leaving the cloud thickness 
as a free parameter or by finding a constraint on the thickness that reproduces 
the observed \feii$_{25.99\micron}$ line. 

In \cloudy\ model calculations, we have assumed the hydrogen density almost zero. 
We now examine if hydrogen is indeed depleted in IRS1. 
In Figure~\ref{fig:sed}, we present a model that contains some amount of 
hydrogen with $n({\rm H})=10$~cm$^{-3}$. The MIR spectrum of this model 
is identical to the spectrum of the reference model as long as the total 
amount of metals and grains are retained the same, but the model with 
hydrogen shows a group of emission lines from hydrogen in optical and NIR
which have not been observed in IRS1. 
We searched for the H$\alpha$ emission around IRS1  
from the VST Photometric H$\alpha$ Survey of the Southern 
Galactic Plane and Bulge \citep[VPHAS+;][]{drew14}\footnote{http://www.vphasplus.org}. 
The predicted H$\alpha$ line flux of the \cloudy\ models are 
$8.9 \times 10^{-21}$ and $8.9 \times 10^{-15}$~\ergs\ 
for the models with $n({\rm H})=10^{-4.9}$ and 10~cm$^{-3}$, respectively,
after applying the extinction by $N_{\rm H}=9.5 \times 10^{21}$~cm$^{-2}$ \citep{gaensler02,koo11}. 
For comparison, the 5$\sigma$ limiting magnitude of VPHAS+ is 
about 20~mag \citep{drew14}, or $1.84 \times 10^{-15}$~\ergs\ in H$\alpha$. 
If IRS1 contains hydrogen, even a small amount of $\lesssim 10$~cm$^{-3}$, 
the H$\alpha$ emission is expected to be detected in the VPHAS+ images,
but no emission has been found around IRS1. This supports our assumption 
that hydrogen is depleted in IRS1.

\section{Crystalline Silicate in \msh}\label{sec:discuss}

\msh\ is the first SNR in which crystalline silicate is observed.
Our analysis in this paper indicates that the elemental abundance of 
the IR compact source IRS1 where crystalline silicate has been detected 
is close to that of SN ejecta with depleted hydrogen and high abundance of metals, 
particularly neon, argon, and iron. 
This implies that IRS1 (and probably IRAS 15099-5856 as well) originates 
from the SN ejecta rather than the mass loss of the SN progenitor
as has been proposed by \citet{koo11}. If this is true, \msh, 
besides the existence of crystalline silicate, is 
a unique object where we can directly observe dust newly formed 
in the ejecta of SNe Ib/c. Observing the pristine SN dust which has not yet been 
processed by the SN reverse shock is also very rare only with a few other 
cases of G54.1+0.3 \citep{temim17} and Cas A \citep{sibthorpe10,delooze17}. 
While FIR observations (e.g., AKARI, Herschel, and ISO) 
have revealed cold dust inside SNRs, 
the spatial resolutions are not high enough to disentangle the SN dust 
from the surrounding ISM dust and examine dust properties 
in detail \citep[e.g.,][and references therein]{sibthorpe10,koo16,chawner20,millard21,rho23}.
The cold and warm dust has been detected in \msh\ as well.
\citet{millard21} estimated 0.03--0.06~\msol\ of warm (46--52~K) and 
4--15~\msol\ of cold (17--20~K) dust, assuming the distance of 5.2~kpc to \msh, 
from the two-component blackbody model fitting of the MIR to FIR spectrum 
obtained by the Long Wavelength Spectrometer (LWS) 
on board the ISO.
Since their spectrum is not background-subtracted, they suggest that the warm and 
cold dust originate from the SN ejecta and background ISM, respectively. 
The dust mass of 0.03--0.06~\msol\ at 5.2~kpc is scaled to 
0.013--0.026~\msol\ at 3.4~kpc. This is several to ten times larger 
than our results, but a direct comparison of the two dust masses is inappropriate 
because the LWS spectrum of \msh\ with a large beam size 
of $\sim$80$\arcsec$ \citep{gry03} includes not only IRS1 but the surrounding, diffuse 
emission.

Dust formation in the SN ejecta of Type II SNe (SNe II) has been widely studied 
by theoretical calculations \citep[e.g.,][]{sarangi13,sarangi15,brooker22} as well as 
the observations of young SNRs, e.g., Cas A \citep{delooze17}, SN 1987A \citep{matsuura15}, 
G54.1+0.3 \citep{rho18}, and the Crab Nebula \citep{gomez12}. 
In contrast, few studies thus far have been carried out for dust condensation 
in the ejecta of SNe Ib/c. 
Observational signatures of dust formation in the SN Ib/c ejecta have been discovered 
only for a few SNe in the nebula phase:
SN Ib 1990I \citep{elmhamdi04}, SN Ib(n) 2006jc \citep{dicarlo08,smith08}, 
SN Ic 2020oi \citep{rho21}, and SN Ic 2021krf \citep{ravi23}.
The molecules which become dust seeds in the SN ejecta and their chemical reactions 
would not be very different between SNe II and Ib/c, but the environments 
in which dust condensation occurs 
would differ depending on SN types and progenitor stars. 
For example, even in the same SN type of IIP,
the amount of dust formed in the ejecta and the degree of grain growth predicted 
by dust condensation models highly depend on the conditions of SN explosion 
such as progenitor mass, explosion energy, mass of $^{56}$Ni, or 
clumpy structure of ejecta \citep{sarangi13,sarangi15,brooker22}.
In the ejecta of SNe IIP, dust starts to form a few hundreds days after 
SN explosion \citep[][and references therein]{sarangi18}. The observations of dust signatures 
in the SN Ib/c ejecta listed above indicate that the onset of dust formation occurs only 50--70 days 
after explosion \citep{dicarlo08,smith08,rho21,ravi23}, with the exception of 
SN 1990I \citep[$\sim$230 days;][]{elmhamdi04}. 
The reason for the early dust condensation in the ejecta of SNe Ib/c is thought to be 
due to the rapid decrease in the gas temperature \citep{nozawa08}. 
For SNe Ib/c, the ejected masses are smaller and the expansion velocities are higher than SNe II 
because the SN Ib/c progenitors have lost most of the hydrogen/helium envelopes before explosion. 
This leads to a lower density of gas in the ejecta, and gas temperature drops down more quickly than 
those in typical SNe II. \citet{nozawa08} calculated dust formation in the SN 2006jc applying 
the SN Ib model of a relatively low-mass (6.9~\msol) helium star progenitor with an ejecta mass 
of 4.9~\msol. Their calculation predicts the gas temperature reaching a typical dust condensation 
temperature ranges of 1,000--2,000~K between 50 and 200 days after the SN explosion, which means 
that dust forms much closer to the explosion center in the ejecta of SNe Ib/c than 
in the SNe II ejecta \citep[Figure~1 of][]{nozawa08}.
A similar process of dust formation might have occurred in \msh\ of which progenitor is also 
speculated as a low-mass helium star.

The different environmental conditions may bring distinctive characteristics of dust formed 
in the SN Ib/c ejecta, for example, the formation of crystalline silicate or 
crystallization of amorphous silicate.
If silicates are formed at high ($>$1,000~K) temperature, the crystalline lattice structure is 
the most favorable state \citep{molster05}. 
In a clumpy ejecta, stoichiometric silicate (Mg$_{2}$SiO$_{4}$) is predicted to be formed 
at high densities \citep{sarangi15}. 
Therefore, the condensation of crystalline silicates could take place in the SN ejecta. 
Alternatively, silicates first formed in amorphous structure could be crystallized 
if there is a high-energy process such as heating by pulsar wind nebula,
although the condensation temperature and condensing phase of dust are 
not simply determined but associated with several factors such as 
gas pressure or gas kinematics \citep{nagahara09,gail13}. 
Then, assuming that crystalline silicates can form in SN dust, 
why have crystalline silicates not been found in other SNRs except \msh? 
This is probably because it is difficult to observe unshocked SN dust 
without a heating source such as Muzzio 10. 
Crystalline silicates, even if present, can be undetected in FIR because 
they exhibit very weak or no spectral signatures at wavelengths longer 
than 40~\micron\ except for the 69~\micron\ feature \citep{koike03,sturm13}. 
We note that G54.1+0.3 is another SNR of which unshocked SN dust has been 
revealed by stellar sources \citep{temim17}. In G54.1+0.3, the newly formed dust is heated 
by massive stars in the SNR which are thought to be the members of the same cluster 
with the SN progenitor. While at least 0.3~\msol\ of dust has been found in G54.1+0.3, 
crystalline silicate dust has not been detected. However, G54.1+0.3 is 
the remnant of an SN IIP, and the environment of dust condensation 
would be different from that of \msh. 

While our results suggest a possibility that the crystalline silicate of IRS1 originates from 
the SN ejecta, the current observational data with limited spatial and 
spectral resolution can neither confirm the ejecta origin 
nor rule out the progenitor origin.
The Gemini/T-ReCS images show a slightly different spatial distribution 
between \neii\ 12.81~\micron\ and Qa 18.30~\micron\ on a spatial 
scale less than one arcsecond (Figure~\ref{fig:trecs}). This is likely because 
the \neii\ line does not trace the Mg-silicate but dust with a very smooth spectrum 
such as Al$_{2}$O$_{3}$ \citep{arendt14}. 
The spatial distributions of various ionic lines will be required to examine 
a correlation between the gas and dust features associated with crystalline silicate. 
The Spitzer IRS spectrum of IRS1 did not resolve the velocity of 
the gas component. Previously, \citet{koo11} favored the progenitor origin 
of IRS1 based on the low ($-160 \pm 560$~\kms) central velocity of the \neii\ line. 
The velocities of the other lines except \feii\ 25.99~\micron\ are similarly 
a few hundreds~\kms\ with an average of $-444$~\kms, 
but the uncertainties are huge, as for \neii. 
The \feii\ line exceptionally shows a large velocity of $-1710 \pm 280$~\kms.
This may imply a different origin of \feii\ from the other lines, but it can be due to 
an inaccurate measurement of the velocity since the \feii\ line is weak and 
embedded between strong dust features (Figure~\ref{fig:irsspec}).

To confirm the origin of the crystalline silicate in \msh\ and explore the possibility 
of the formation of crystalline silicates in SN ejecta, further observations 
as well as theoretical investigations are required. Particularly, it is crucial to examine 
the spatial distributions of gas and dust in IRAS 15099-5856 
through the MIR observations with high resolution and sensitivity (e.g., JWST/MIRI). 
If the origin of the crystalline silicate is found to be the SN ejecta, 
\msh\ will give an unprecedented opportunity to investigate dust formation in 
SN ejecta, which is not yet clearly known.

\section{Summary and Conclusion}\label{sec:summary}

We have presented the MIR imaging observations and analysis of 
the compact IR source IRS1 of IRAS 15099-5856 in the SNR \msh, which is 
the first and only object with crystalline silicate dust associated with SNRs so far. 
The MIR images obtained by using Gemini/T-ReCS revealed the morphology of IRS1 and spatial 
distributions of gas and dust at a spatial resolution of $\lesssim$ 1$\arcsec$.
We have also presented the analysis of the Spitzer IRS spectrum of IRS1 
that was previously investigated with the models of thermal emission from 
multiple independent dust components \citep{koo11}. In this paper, 
we have analyzed the ionic lines and modeled the spectrum considering the geometry 
and energy balance to derive the chemical abundance of gas as well as dust parameters. 
The derived abundance is close to that of SN ejecta 
with poor hydrogen and enhanced metals. This suggests the ejecta origin for 
the crystalline silicate and 
may imply the possibility of the formation of crystalline silicate 
in SN ejecta, but the current observational data are still limited in 
spatial and spectral resolution. If the origin of the crystalline silicate 
in IRAS 15099-5856 is confirmed as the SN ejecta by future observations, 
\msh\ will be a unique, invaluable object that proves the formation of 
crystalline silicate in SN ejecta and where we can directly observe 
newly-formed dust in the ejecta of SNe Ib/c. 
In the following, we summarize our main results.

\begin{enumerate}

\item The Gemini/T-ReCS images show a complicated, extended morphology of IRS1 
with bright clumps and diffuse emission in \neii\ 12.81~\micron\ and 
Qa 18.30~\micron. The \neii cont image with no emission and the Si-6 image 
with the almost same emission as \neii\ indicate that there is no other line or 
strong continuum emission. 
The T-ReCS images confirm the previous prediction \citep{koo11} that 
IRS1 is extended and externally heated by the nearby O star Muzzio 10. 
We estimated the \neii\ and Qa flux of IRS1 from the T-ReCS images and 
compared them with the flux derived from the Spitzer IRS spectrum. 

\item The Spitzer spectrum of IRS1 shows prominent dust features at 23, 27, and 
34~\micron\ that can be explained by crystalline silicate dust. 
We also detected several ionic lines of \ariii\ 8.99~\micron, \siv\ 10.51~\micron,
\neii\ 12.81~\micron, \neiii\ 15.56~\micron, \siii\ 18.71~\micron, 
and \oiv\ 25.89/\feii\ 25.99~\micron. The \oiv\ and \feii\ are not resolved at 
the spectral resolving power of the Spitzer IRS LL module, but it is likely \feii.
The estimated line flux is from 0.13 to 8.21 $\times 10^{-12}$~\ergs. The line widths 
are comparable to the spectral resolving power, i.e., the velocity is not resolved.

\item We compared the line ratios of \neiii$_{15.56\micron}$/\neii$_{12.81\micron}$ 
versus \siv$_{10.51\micron}$/\siii$_{18.71\micron}$ of various astronomical objects 
with the observed line ratios of IRS1 on a model grid generated 
by \cloudy\ \citep{ferland13}. The line ratio diagram shows that the abundance 
of IRS1 is rather close to the nova abundance with enhanced neon. 
The absence of hydrogen lines in the Spitzer spectrum further suggests that 
hydrogen is depleted in IRS1.

\item We modeled the Spitzer spectrum of IRS1 using 
the photoionization code \cloudy\ \citep{ferland13}. We assumed 
the cloud IRS1 externally heated by an O4.5III star (Muzzio 10) separated by 0.45~pc. 
For the gas, the nova abundance was initially adopted but with hydrogen depleted. 
For the dust species, crystalline olivine (Mg$_{1.9}$Fe$_{0.1}$SiO$_{4}$), FeO, 
Mg$_{0.6}$Fe$_{0.4}$O, and amorphous silicate were included. Spherical 
dust grains of 0.25~\micron\ (FeO) and 0.1~\micron\ (the others) were assumed.
We first fitted the 
dust features and adjusted the abundance of nitrogen, oxygen, neon, sulphur, 
argon, and iron to find a model that reproduces the observed lines. 
We have derived a reference model that fairly well fits the Spitzer spectrum 
and discussed the factors that affect the models.

\item The reference model fits the dust features at 27 and 34~\micron, while it does 
not well fit the 23~\micron\ feature and steeply-increasing continuum between 
15 and 20~\micron. The derived dust mass is $2.02 \times 10^{-3}$~\msol\ 
in total, mostly contributed by amorphous silicate. 
The dust temperature is between 67 and 79~K. 
The total dust mass is about 31\% of the mass 
previously derived by multi-component blackbody fitting \citep{koo11}, 
but the dust mass depends on the separation to the heating source which was 
fixed in our modeling. 
The reference model predicts the emission lines with comparable 
intensities to the observation except the \siv\ 10.51\micron\ and
\feii\ 25.99~\micron\ lines. The difference in \siv$_{10.51\micron}$ is 
likely due to the overestimation of the observed line flux. On the other hand, 
\feii$_{25.99\micron}$ is predicted much weaker (15\%) than 
the observed intensity. This can be due to the overestimated 
flux of the observed line in part, but we have also found that the ionization 
fraction of iron is sensitive to the cloud thickness which was also fixed in our modeling. 
The model can be improved by changing dust properties (shape or size) 
and geometrical parameters (e.g, cloud thickness or the separation between 
IRS1 and Muzzio 10), although it is beyond the scope of this paper. 

\item The \cloudy\ modeling of the Spitzer spectrum indicates that 
the elemental abundance of IRS1 is close to that of SN ejecta 
with depleted hydrogen and high abundance of metals, particularly 
neon, argon, and iron. This implies that IRS1 (and IRAS 15099-5856) 
is the SN ejecta and that crystalline silicate can form 
in SN ejecta. To confirm this, further observations to resolve 
the spatial distributions of various ionic lines, dust continuum, and 
crystalline silicate will be required. 

\end{enumerate}

\begin{acknowledgments}
This work is based in part on observations obtained at the international Gemini Observatory, a program of NSF’s NOIRLab, which is managed by the Association of Universities for Research in Astronomy (AURA) under a cooperative agreement with the National Science Foundation on behalf of the Gemini Observatory partnership: the National Science Foundation (United States), National Research Council (Canada), Agencia Nacional de Investigaci\'{o}n y Desarrollo (Chile), Ministerio de Ciencia, Tecnolog\'{i}a e Innovaci\'{o}n (Argentina), Minist\'{e}rio da Ci\^{e}ncia, Tecnologia, Inova\c{c}\~{o}es e Comunica\c{c}\~{o}es (Brazil), and Korea Astronomy and Space Science Institute (Republic of Korea).
The Gemini program identification number associated with the data is GS-2012A-C-4.
This work is based in part on observations made with the Spitzer Space Telescope, which was operated by the Jet Propulsion Laboratory, California Institute of Technology under a contract with NASA.
This work has made use of data from the European Space Agency (ESA) mission
{\it Gaia} (\url{https://www.cosmos.esa.int/gaia}), processed by the {\it Gaia}
Data Processing and Analysis Consortium (DPAC,
\url{https://www.cosmos.esa.int/web/gaia/dpac/consortium}). Funding for the DPAC
has been provided by national institutions, in particular the institutions
participating in the {\it Gaia} Multilateral Agreement.
This research of H.-J. K. was supported by Basic Science Research Program through the National Research Foundation of Korea (NRF) funded by the Ministry of Education (RS-2023-00246733). This research of B.-C. K. was supported by the Basic Science Research Program through the National Research Foundation of Korea (NRF) funded by the Ministry of Science, ICT and Future Planning (2020R1A2B5B01001994, RS-2023-00277370). T.O. is supported by the Japan Society for the Promotion of Science (JSPS) KAKENHI Grant Numbers JP 23K22532 and 24K07087.

\end{acknowledgments}

\facilities{Gemini:South (T-ReCS), Spitzer  (IRS)}
\software{\cloudy\ version C13~\citep{ferland13}, MPFIT~\citep{markwardt09} }

\end{document}